\documentclass[12pt]{article}
\usepackage{amssymb}
\usepackage{amsfonts}
\usepackage{amsmath}
\usepackage{graphicx}
\usepackage{color}
\usepackage{rotating}

\def\Z{\mathbb Z}

\def\R{\mathbb R}
\def\NN{\mathbb N}

\newcommand{\ve}{\varepsilon}
\renewcommand{\th}{\theta}

\newcommand{\De}{{\Delta\vphantom{\big|}}}
\newcommand{\Om}{\Omega}

\newcommand{\E}{{\scriptscriptstyle E}}
\newcommand{\wb}{{\bar{w}}}
\newcommand{\rb}{{\bar{\rho}}}
\newcommand{\cD}{{\cal D}}
\newcommand{\cH}{{\cal H}}
\newcommand{\cL}{{\cal L}}
\newcommand{\cM}{{\cal M}}

\newcommand{\s}{{\mathrm{s}}}
\renewcommand{\c}{{\mathrm{c}}}

\newcommand{\+}{{\dagger}}
\newcommand{\sfrac}[2]{{\textstyle\frac{#1}{#2}}}
\newcommand{\half}{{\sfrac12}}
\newcommand{\pa}{\partial}

\newcommand{\res}{{\mathrm{res}}}

\newcommand{\im}{{\mathrm{i}}}
\newcommand{\ep}{{\mathrm{e}}}

\newcommand{\beq}{\begin{equation}}
\newcommand{\eeq}{\end{equation}}
\newcommand{\eq}{\end{equation}}
\newcommand{\bea}{\begin{eqnarray}}
\newcommand{\eea}{\end{eqnarray}}
\newcommand{\with}{{\quad{\rm with}\quad}}
\newcommand{\for}{{\quad{\rm for}\quad}}
\renewcommand{\and}{{\quad{\rm and}\quad}}
\newcommand{\und}{{\qquad{\rm and}\qquad}}
\renewcommand{\=}{\ =\ }

\advance \textheight by \topskip
\makeatletter
\@addtoreset{equation}{section}

\makeatother

\begin{document}

\begin{titlepage}
\setcounter{page}{0}
\begin{flushright}
ITP--UH--14/15\\
\end{flushright}

\vskip 1.5cm

\begin{center}

{\LARGE\bf
The tetrahexahedric angular Calogero model
}

\vspace{12mm}
{\Large Francisco Correa${}^{\+\times}$ and \
Olaf Lechtenfeld${}^{\times}$ 
}
\\[8mm]

\noindent ${}^\+${\em 
Centro de Estudios Cient\'ificos (CECs)\\
Casilla 1468, Valdivia, Chile } \\
{Email: correa@cecs.cl}\\[6mm]

\noindent ${}^\times${\em
Institut f\"ur Theoretische Physik and Riemann Center for Geometry and Physics\\
Leibniz Universit\"at Hannover \\
Appelstra\ss{}e 2, 30167 Hannover, Germany }\\
{Email: lechtenf@itp.uni-hannover.de}\\[6mm]

\vspace{12mm}

\begin{abstract}
\noindent
The spherical reduction of the rational Calogero model 
(of type $A_{n-1}$ and after removing the center of mass)
is considered as a maximally superintegrable quantum system,
which describes a particle on the $(n{-}2)$-sphere subject
to a very particular potential. We present a detailed analysis
of the simplest non-separable case, $n{=}4$, whose potential
is singular at the edges of a spherical tetrahexahedron. 
A complete set of independent conserved charges and of Hamiltonian 
intertwiners is constructed, and their algebra is elucidated.
They arise from the ring of polynomials in Dunkl-deformed
angular momenta, by classifying the subspaces invariant
and antiinvariant under all Weyl reflections, respectively.
\end{abstract}

\end{center}

\end{titlepage}

\section{Introduction and summary}

\noindent
The Calogero (or Calogero--Moser) model~\cite{Cal69,Cal71,CalMar74}\footnote{
For reviews, see e.g.~\cite{OlshaPere81-rev,OlshaPere83-rev,Poly1,Poly06-rev}.}
is the paradigmatical $n$-particle integrable system in one space dimension.
Originally defined for the root system of $A_1\oplus A_{n-1}$, the Calogero model 
was quickly generalized for any finite Coxeter group of rank~$n$~\cite{OlshaPere81-rev}.
In particular the rational version is remarkable for its conformal properties 
and its maximal superintegrability~\cite{Woj,Kuz} (see also \cite{BarReg77}).
Moreover, when the coupling constant~$g$ is integral, the quantum model enjoys 
additional and algebraically independent conserved quantities, which make it
`analytically integrable'~\cite{ChaVes90,Chalykh96}. In this case, intertwining operators
relate the energy spectrum with that of the free theory (at $g{=}0$ or $g{=}1$).
This structure was worked out in the 90s by a Russian group~\cite{ChaVes90,Chalykh96}
using Darboux dressing and by Dutch mathematicians~\cite{Opdam,Heckman} employing
Dunkl operators~\cite{Dunkl89,DunklXu}.

Maximal superintegrability for a quantum system of phase-space dimension~$2n$
has been characterized in the literature; for a recent account see~\cite{Resh}
where it is termed `maximal degenerate integrability'.
The Hamiltonian $H$ is part of a rank-$n$ abelian algebra of Liouville charges
(commuting Hamiltonians), but this is embedded in a larger commutant of~$H$, which
is spanned by the Liouville charges plus $n{-}1$ additional integrals of motion
(Wojciechowski charges). The latter do not commute with one another but form a
polynomial algebra~\cite{Kuz} together with the Liouville charges.
Denoting the commutant generators by $C$ and distinguishing
between Liouville charges~$C'$ and Wojciechowski charges $C''$ so that
$\{C\}=\{C',C''\}$, one has (see, e.g.~\cite{CoLePl13})
\begin{equation} \label{alg}
[C',C']\=0 \quad,\qquad
[C'',C']\={\cal A}(C',C'') \quad,\qquad
[C'',C'']\={\cal B}(C',C'')\ ,
\end{equation}
where ${\cal A}$ and ${\cal B}$ are polynomial in $C'$ and linear in~$C''$,
and we have suppressed the counting labels.
This implies that each $H$~eigenspace splits into simultaneous 
eigenspaces of the other $n{-}1$ Liouville charges. 
Hence, we can find an $H$~eigenbasis labelled by
$n$ quantum numbers $(\ell_\alpha)$ with $\alpha=1,\ldots,n$.
Furthermore, these simultaneous eigenspaces carry a representation of the 
algebra~(\ref{alg}), which governs their degeneracies.

Since the rational Calogero Hamiltonian is part of an SL(2,$\R$) conformal algebra, 
it is natural to reduce the rational model to an integrable system on a sphere.
Restricting ourselves to the customary $A_1\oplus A_{n-1}$ model, 
one may employ the translational invariance and firstly separate the $A_1$ part 
representing the center-of-mass motion. The relative-motion Hamiltonian may
be interpreted as ruling a single particle in $\R^{n-1}$, subject to a
potential which is singular at the reflection hyperplanes and decays like $r^{-2}$
with distance~$r$ from the origin. Secondly, this radial scaling suggests a
further codimension-one reduction to $S^{n-2}$, which defines the
`angular (relative) rational Calogero model'.\footnote{
It is investigated already in the appendices of~\cite{Cal71}
(see also \cite{Cal69,CalMar74} for $n{=}3$).}
This system describes a particle moving on $S^{n-2}$ in a collection of
$\frac12n(n{-}1)$ Higgs-oscillator potentials~\cite{higgs,leemon}, 
each one centered at a positive root of $A_{n-1}$ (and its antipode) and blowing up
at the intersection of the reflection hyperplane with the unit sphere.
These singular loci tessellate the $(n{-}1)$-sphere into $n!$ hyperspherical simplices
in which the particle is trapped.

A number of classical features of angular rational Calogero model 
(center-of-mass-reduced or not) have been recently analyzed by an Armenian 
group~\cite{HNY,HKLN,LNY,HLNS,HLN,HLNSY}.
Intertwining relations for the quantum model were investigated by M.~Feigin
already in~\cite{Feigin03}, but more recently the energy spectra and eigenstates
have been constructed~\cite{FeLePo13}, and the algebra of angular Dunkl operators
has been studied~\cite{FeHa14}. These partial results leave open the following questions:
\begin{itemize}
\addtolength{\itemsep}{-6pt}
\item What is a complete set of algebraically independent conserved charges?
\item Can one construct a subset of Liouville charges (in mutual involution)?
\item What is a complete set of algebraically independent Hamiltonian intertwiners?
\item Can one find additional conserved quantities which get intertwined?
\item What kind of algebra is generated by the conserved charges and intertwiners?
\item Are these models maximally quantum super- or even analytically integrable?
\end{itemize}
The answers are not easily obtained for the angular Calogero models because
they are linked to nontrivial mathematical problems. For instance, the
commutant of~$H$ is the ring of all Weyl-invariant polynomials in the angular Dunkl
operators (for a definition see the following section), and it is not obvious 
how to construct a minimal set of algebraically independent generators. 
Moreover, to establish the superintegrability one should find the maximal abelian
subalgebras (i.e.\ the Liouville charges) in this ring.

Lacking a general strategy to attack these issues, we take an explicit look at the
simplest examples, $n{=}3$ and $n{=}4$. The first one, leading to a rank-one
angular system, is quite trivial but well known, while the second one is still tractable
and serves to illustrate most of the general concepts.
Therefore, in this paper, we exemplarily analyze the rank-two quantum system which 
is obtained from the four-particle rational Calogero model, via
translational reduction to $\R^3$ and then radial reduction to $S^2$. It describes
a particle on the two-sphere trapped in one of 24 right isosceles spherical triangles
which make up the spherical projection of a so-called tetrahexahedron.

After presenting general (and mostly previous) results on the $A_{n-1}$ angular models
in Section~2, the essentially trivial $n{=}3$ case, known as the P\"oschl-Teller model, 
is reworked in Section~3 as a warmup example. Section~4 introduces the $n{=}4$ model
and recapitulates its energy spectrum and eigenstates~\cite{FeLePo13}, before our
results concerning conserved charges and intertwiners are collected in Section~5.
There, we establish that any conserved charge of the tetrahexahedric angular
Calogero model is built from two independent operators, say $J_4$ and $J_6$ 
(besides the Hamiltonian), and that this system features two independent 
Hamiltonian intertwiners, which we denote by $M_3$ and $M_6$.
We provide explicit expressions for both and work out their (non-standard)
intertwining relations with the elementary conserved charges. 
Since one cannot produce two commuting words with $J_4$ and $J_6$, the model
enjoys only two charges in involution but, 
given three independent conserved quantities, it is maximally superintegrable.
For integer value of the coupling, an additional independent conserved charge 
introduces a $\Z_2$ grading (in the spirit of ‘nonlinear supersymmetry’ without 
fermion-like degrees of freedom~\cite{bososusy}) and renders the model
analytically integrable.
Various interesting algebraic relations are collected in the Appendix.

\section{The angular (relative) rational Calogero model}

\noindent
We parametrize the $n$-particle quantum phase space with coordinates $x^\mu$ and
momenta $p_\nu$, subject to the canonical commutation relations (setting $\hbar=1$)
\begin{equation}
[\,x^\mu\,,\,p_\nu\,] \= \im\,\delta^\mu_{\ \nu} \quad\with \mu,\nu=1,\ldots,n\ .
\end{equation}
The center-of-mass coordinate and momentum,
\begin{equation}
X\=\sfrac1n\sum_\mu x^\mu \und P\=\sum_\nu p_\nu \qquad\Rightarrow\qquad  [X,P]=\im \ ,
\end{equation}
decouple in the rational Calogero Hamiltonian
\begin{equation}
H_{\textrm{cal}} 
\= \sfrac12 \sum_\nu p_\nu^2\ +\ \sum_{\mu<\nu} \frac{g(g{-}1)}{(x^\mu{-}x^\nu)^2} 
\= \sfrac{1}{2n} P^2\ +\ H
\end{equation}
and will be ignored in our starting Hamiltonian
\begin{equation}
H \= \sum_{\mu<\nu} \Bigl\{ \sfrac{1}{2n}(p_\mu{-}p_\nu)^2\ +\ 
\frac{g(g{-}1)}{(x^\mu{-}x^\nu)^2} \Bigr\}
\end{equation}
which describes the relative motion of ($A_{n-1}$) Calogero particles on the real line,
or of a single quantum particle in $\R^{n-1}$ subject to a potential singular at the $A_{n-1}$
Weyl chamber walls. The Hamiltonian is invariant under the reflection $g\to1{-}g$
of the real coupling~$g$, but higher conserved charges and intertwiners will not share this property.

Since this Hamiltonian is scale covariant, it pays to introduce on $\R^{n-1}$
polar coordinates $(r,\vec\th)$ with a (relative) radius $r$ and $n{-}2$ angles $\vec\th$
together with their momenta $p_r$ and $L_{\vec\th}$,  
\begin{equation}
\sfrac1n\sum_{\mu<\nu} (x^\mu{-}x^\nu)^2 \= r^2 \und
\sfrac1n\sum_{\mu<\nu} (p_\mu{-}p_\nu)^2 \= 
p_r^2 + \sfrac{1}{r^2}L^2 + \sfrac{(n-2)(n-4)}{4\,r^2} \ .
\end{equation}
The (relative) angular momentum squared $L^2$ is best expressed by parametrizing
the relative particle motion in $\R^{n-1}$ with coordinates $y^i$ where $i=1,\ldots,n{-}1$,
\begin{equation}
r^2 \= \sum_{i=1}^{n-1} (y^i)^2 \quad,\qquad p_i\equiv p_{y^i} \quad,\qquad
L_{ij} \= -\im(y^i p_j-y^j p_i) \quad,\qquad L^2 \= -\sum_{i<j}L_{ij}^2\ .
\end{equation}
Jacobi coordinates are a suitable choice for $\{y^i\}$.
Together with a dilatation generator~$D$ and a special conformal generator~$K$, 
the Hamiltonian $H$ forms an SL(2,$\R$) commutator algebra, 
as is easily seen in the polar representation
\begin{equation}
H \= \sfrac12 p_r^2\ +\ \sfrac{(n-2)(n-4)}{8\,r^2}\ +\ \sfrac{1}{r^2} H_\Om \quad,\qquad
D \= \half(r\,p_r+p_r r) \quad,\qquad K \= \half r^2\ .
\end{equation}
Here, the angular (relative) Hamiltonian $H_\Om$ is independent of the radial coordinate
or momentum, and it coincides (up to a constant shift) with the Casimir operator of the
conformal SL(2,$\R$),
\begin{equation}
H_\Om \= \sfrac12 L^2  + U(\vec\th)
\= C - \sfrac18 (n{-}1)(n{-}5)
\qquad\textrm{with}\qquad  
C \= K\,H + H\,K - \half D^2\ .
\end{equation}
This defines our angular Calogero system (for $A_{n-1}$), which describes a quantum particle
on the sphere $S^{n-2}$ under the influence of the potential
\begin{equation}
U(\vec\th) \= r^2 \sum_{\mu<\nu} \frac{g(g{-}1)}{(x^\mu{-}x^\nu)^2} 
\ \=\ r^2 \sum_{\alpha\in{\cal R}_+} \frac{g(g{-}1)}{(\alpha\cdot y)^2}
\ \=\ \sfrac{g(g{-}1)}{2} \sum_{\alpha\in{\cal R}_+} \cos^{-2}\th_\alpha \ ,
\end{equation}
where ${\cal R}_+$ is the set of positive roots for $A_{n-1}$ and $\th_\alpha$ denotes
the angle (geodetic arc length on $S^{n-2}$) between the point $\vec\th$ and 
the intersection of the ray in direction~$\alpha$ with the unit sphere.
Each of the $\sfrac12n(n{-}1)$ contributions of $\cos^{-2}$ form 
is also known as a `Higgs oscillator'~\cite{higgs,leemon}.
This potential is a very special one as it is tied to the root system.
Its singular walls are of codimension one and tesselate the sphere 
into $n!$ hyperspherical simplices.

For the spectral problem, let us pass to the position representation 
and encode momenta via differental operators,
\begin{equation}
p_i\ \mapsto\ -\im\pa_i \qquad\Longrightarrow\qquad
p_r\ \mapsto\ -\im\bigl(\pa_r + \sfrac{n-2}{2\,r}\bigr)
\end{equation}
so that
\begin{equation}
\begin{aligned} \label{Hdiff}
H\ &\mapsto\ -\sfrac12 \bigl(\pa_r^2 + \sfrac{n-2}{r}\pa_r\bigr)\ +\ \sfrac{1}{r^2} H_\Om 
\qquad\Leftrightarrow \\[4pt]
H_\Om\ &\mapsto\ r^2 H\ +\ \sfrac12(r\pa_r+n{-}3)\,r\pa_r \ .
\end{aligned}
\end{equation}
It is customary to remove the first-order radial derivative term via a similarity transformation,
\begin{equation} \label{sim}
\Psi(r,\vec\th) \= r^{-\frac{n-2}{2}} \, u(r,\vec\th) 
\und
H \= r^{-\frac{n-2}{2}} H'\,r^{\frac{n-2}{2}}\ ,
\end{equation}
so that the Hamiltonian on $\R^{n-1}$ acts as follows,
\begin{equation}
H'\,u\ \mapsto\ -\sfrac12 \bigl(\pa_r^2 - \sfrac{(n-2)(n-4)}{4\,r^2}\bigr)\,u\ +\ \sfrac{1}{r^2} H_\Om\,u\ .
\end{equation}
As the potential is repulsive, the spectrum is positive semi-definite and continuous,
\begin{equation}
H\,\Psi_{\E,q} \= E\,\Psi_{\E,q} \qquad\Longleftrightarrow\qquad
H'\,u_{\E,q} \= E\,u_{\E,q} 
\qquad\textrm{with}\qquad E\in\R_{\ge0} \ ,
\end{equation}
where the wave function depends on further (suppressed) quantum numbers, 
but $q$ parametrizes the angular Hamiltonian eigenvalues~\cite{FeLePo13} 
(see also the appendices of~\cite{Cal71}),
\begin{equation} \label{HOmspec}
H_\Om\,v_q \= \ve_q\,v_q \quad\with\quad \ve_q \= \half q\,(q+n-3) \and q\in\R_{\ge0}\ .
\end{equation}
For vanishing potential ($g{=}0$ or $1$), $H_\Om=\sfrac12L^2$, and $q=\ell\in\NN_0$ is the familiar
total angular momentum for a free particle on $S^{n-2}$, with a degeneracy of 
$\bigl( \begin{smallmatrix} \ell{+}n{-}2 \\ n{-}2 \end{smallmatrix} \bigr)
-\bigl( \begin{smallmatrix} \ell{+}n{-}4 \\ n{-}2 \end{smallmatrix} \bigr)$.
The interaction greatly reduces the degeneracy and shifts this quantum number~\cite{FeLePo13},
\begin{equation} \label{q}
q \= \half n(n{-}1)g+\ell \und 
\ell\=3\ell_3+4\ell_4+\ldots+n\ell_n
\with \ell_\mu\in\NN_0\ ,
\end{equation}
where the decomposition $\ell\to(\ell_3,\ell_4,\ldots,\ell_n)$ into additional quantum numbers
yields the correct degeneracy,
\begin{equation} \label{deg}
\textrm{deg}_n(\ve_q) \= p_n(\ell)-p_n(\ell{-}1)-p_n(\ell{-}2)+p_n(\ell{-}3)\ ,
\end{equation}
with $p_n(\ell)$ denoting the number of partitions of $\ell$ into integers not larger than~$n$ or,
equivalently, the number of partitions of $\ell$ into at most $n$~integers. Its generating function reads
\begin{equation}
p_n(t)\ :=\ \sum_{\ell=0}^\infty p_n(\ell)\,t^\ell \= \prod_{m=1}^n \bigl(1-t^m\bigr)^{-1}\ ,
\end{equation}
and the large-$\ell$ asymptotics is $p_n(\ell)=\frac{\ell^{n-1}}{n!\,(n{-}1)!}+O(\ell^{n-2})$, implying
$\textrm{deg}_n(\ve_q)\sim\ell^{n-2}$ for $\ell\to\infty$. 
Note that the degeneracy does not depend on~$g$ but only on $\ell$ (for any fixed $n$).
Up to $n{=}5$, fairly explicit formulae may be found~\cite{partitions}:
\begin{equation} \label{degsmall}
\begin{aligned}
\textrm{deg}_3(\ell) &\= \begin{cases}
0 &\for \ell=1,2 \ \textrm{mod} \ 3 \\ 
1 &\for \ell=0 \ \textrm{mod}\ 3
\end{cases} \ ,\\
\textrm{deg}_4(\ell) &\= \Bigl\lfloor\frac{\ell}{12}\Bigr\rfloor \ +\ \begin{cases}
0 &\for \ell=1,2,5\ \textrm{mod} \ 12 \\ 
1 &\for \ell=\textrm{else} \ \textrm{mod} \ 12
\end{cases} \ ,\\
\textrm{deg}_5(\ell) &\= \Bigl\lfloor\frac{6\ell^2+72\ell-89}{720}\Bigr\rfloor \ +\ \begin{cases}
0 &\for \ell=2,22,26,46\ \textrm{mod} \ 60 \\ 
2 &\for \ell=0,48\  \textrm{mod} \ 60 \\
1 &\for \ell=\textrm{else} \ \textrm{mod} \ 60
\end{cases} \ .
\end{aligned}
\end{equation}

The wave functions may also be given explicitly. From
\begin{equation}
\bigl( \pa_r^2 - (q{+}\sfrac{n}2{-}1)(q{+}\sfrac{n}2{-}2)r^{-2} + 2E \bigr)\,u_{\E,q} \= 0
\end{equation}
we read off that
\begin{equation} \label{radfct}
u_{\E,q}(r,\vec\th) \= \sqrt{r}\,J_{q+(n-3)/2}({\scriptstyle\sqrt{2E}}\,r)\,v_q(\vec\th)\ ,
\end{equation}
revealing a radial wave function of Bessel type.
The angular wave function $v_q(\vec\th)$ depends on all additional quantum numbers 
$(\ell_3,\ell_4,\ldots,\ell_n)$ and is a bit harder to find~\cite{FeLePo13}.
It can be constructed in the following way,
\begin{equation} \label{vellg}
v_q(\vec\th) \ \equiv\ v_\ell^{(g)}(\vec\th) \ \sim\ 
r^{n-3+q} \prod_{\mu=3}^n \biggl(\sum_{\nu=1}^n (\cD_\nu)^\mu\biggr)^{\ell_\mu}\,\De^g\,r^{3-n-n(n-1)g}\ ,
\end{equation}
where
\begin{equation}
\De \= \prod_{\mu<\nu} \bigl(x^\mu-x^\nu) \= \prod_{\alpha\in{\cal R}_+} \alpha\cdot y
\end{equation}
is the Vandermonde of the original $\R^n$ coordinates and $\cD_\nu$ denotes the
so-called Dunkl operator~\cite{Dunkl89,DunklXu},
\begin{equation}
\cD_\nu \= \pa_\nu \ -\ g\sum_{\rho(\neq\nu)}\frac{1}{x^\nu{-}x^\rho}\,s_{\nu\rho}
\quad,\qquad [ \cD_\mu\,,\,\cD_\nu ] \= 0\ ,
\end{equation}
with $s_{\nu\rho}$ permuting the position and momenta of the $\nu$th with those of the $\rho$th particle.
The Dunkl operators play a central role in asserting the integrability of the Calogero model,
because deforming $\pa_\nu\to\cD_\nu$ essentially creates the interacting system from the free one.
Their Newton sums yield the Liouville charges of the Calogero system, and their mutual commutativity
guarantees that of the charges. 

It is convenient to switch again to the relative coordinates $\{y^i\}$, 
in which the Dunkl-deformed momenta and angular momenta take the form (up to a factor of $-\im$)
\begin{equation}
\cD_i \= \pa_i \ -\ g\sum_{\alpha\in{\cal R}_+} \frac{\alpha_i}{\alpha\cdot y}\,s_\alpha
\und \cL_{ij} \= -(y^i\cD_j - y^j\cD_i)\ ,
\end{equation}
respectively, where $s_\alpha$ is the Weyl reflection on the hyperplane orthogonal to the root~$\alpha$. 
Polynomials in the Dunkl operators are not just differential operators but also act by Weyl reflections.
Their restriction to functions totally symmetric under the Weyl group is called `residue' (denoted by `res')
and produces a purely differential operator. In this way, the Hamiltonian can be obtained from the
homogeneous symmetric polynomial of degree two,
\begin{equation}
\cH \= -\sfrac{1}{2n} \sum_{\mu<\nu} \bigl(\cD_\mu-\cD_\nu)^2
\= -\half\sum_i \cD_i^2
\qquad\Longrightarrow\qquad
H \= \textrm{res}(\cH)\ .
\end{equation}
Analogously, the symmetric quadratic polynomial in the angular Dunkl operators $\cL_{ij}$ 
shifted by a certain pure-reflection term,
\begin{equation}
\cH_\Om \= -\sfrac12\sum_{i<j}\cL_{ij}^2 \ +\ \sfrac12\,S\,(S+n{-}3)
\qquad\textrm{with}\qquad S \= g\sum_\alpha s_\alpha\ ,
\end{equation}
provides the angular Hamiltonian by taking the residue~\cite{FeHa14},\footnote{
We note that $\cL^2=-\sum_{i<j}\cL_{ij}^2$ is a non-negative operator.}
\begin{equation}
H_\Om \= \textrm{res}(\cH_\Om) \= 
\half\textrm{res}\bigl(\cL^2\bigr)\ +\ 
\half\sfrac{n(n{-}1)}{2}g\,\bigl(\sfrac{n(n{-}1)}{2}g+n{-}3\bigr) \=
\half\textrm{res}\bigl(\cL^2\bigr)\ +\ \ve_q(\ell{=}0)\ .
\end{equation}
We remark that the residue of the pure-reflection term simply produces the ground-state energy
$\ve_q(\ell{=}0)$.
Clearly, $H$ and $H_\Om$ may be considered as `Dunkl deformations' of ($-\sfrac12$ times) the
Laplacian on $\R^{n-1}$ and $S^{n-2}$, respectively.

To express the angular wave function $v_q(\vec\th)$ in terms of the relative coordinates,
we must rewrite $\Delta$ in terms of $\{y^i\}$ and the Newton sums of the $\cD_\nu$ in (\ref{vellg}) 
in terms of the $\cD_i$, which depends on our choice for $\{y^i\}$. 
Phrased differently, we need to identify the totally Weyl-symmetric homogeneous polynomials
$\sigma_\mu\bigl(\{\cD_i\}\bigr)$ of orders $\mu=3,4,\ldots,n$, 
to go inside the large brackets of~(\ref{vellg}),
\begin{equation} \label{vellg2}
v_q(\vec\th) \ \equiv\ v_\ell^{(g)}(\vec\th) \ \sim\ 
r^{n-3+q} \biggl( \prod_{\mu=3}^n \sigma_\mu\bigl(\{\cD_i\}\bigr)^{\ell_\mu} \biggr)\,
\De^g\,r^{3-n-n(n-1)g}\ .
\end{equation}
From the form of~(\ref{vellg2}) one can infer that $v_q(\vec\th)$ is 
a ratio of a degree-$q$ homogeneous polynomial $h_q(y)$ in $\{y^i\}$
to the $q$th power of the radial coordinate~$r=\bigl[\sum_i(y^i)^2\bigr]^{-1/2}$.
From (\ref{Hdiff}), the $r$~independence of $v_q(\vec\th)$ and (\ref{HOmspec}) it follows that
the polynomial $h_q(y)$ is annihilated by~$H$,
\begin{equation}
h_\ell^{(g)}\ :=\ r^q\,v_\ell^{(g)}(\vec\th) \qquad\Longrightarrow\qquad H\,h_\ell^{(g)}\=0\ ,
\end{equation}
and hence may be viewed as a Dunkl-deformed harmonic polynomial on~$R^{n-1}$.
Another property visible from (\ref{vellg2}) is that $h_q(y)$ contains a factor of $\De^g$,
which may be split off to define
\begin{equation}
\widetilde{h}_\ell^{(g)} \ :=\ \De^{-g}\,h_\ell^{(g)}
\qquad\Longrightarrow\qquad \widetilde{H}\,\widetilde{h}_\ell^{(g)}\=0\ ,
\end{equation}
which then is in the kernel of the Hamiltonian `in the potential-free frame',
\begin{equation}
\widetilde{H} \= \De^{-g}\,H\,\De^{g}
\= -\sum_{\mu<\nu} \Bigl\{ \sfrac{1}{2n}(\pa_\mu{-}\pa_\nu)^2\ +\ 
\frac{g}{x^\mu{-}x^\nu}(\pa_\mu{-}\pa_\nu)\Bigr\}\ .
\end{equation}
The potential-free Dunkl-deformed harmonic polynomial $\widetilde{h}_\ell^{(g)}$
is homogeneous of degree~$\ell$ only, for any value of~$g$.
In particular, for the ground state one has
\begin{equation}
\widetilde{h}_0^{(g)} \= 1 \qquad\Longrightarrow\qquad h_0^{(g)} \= \De^g \ ,
\end{equation}
and hence the full ground-state wave function is totally symmetric (antisymmetric)
under particle permutations or Weyl reflections for even (odd) integer values of~$g$. 
Since all other ingredients besides $\De^g$ in~(\ref{vellg}) are completely symmetric,
this symmetry property of the integer-$g$ ground state extends to all excited states above it.
Furthermore, the reflection symmetry $g{+}1\leftrightarrow{-}g$ is broken since one tower
of states is Weyl symmetric while the other one is antisymmetric.

It is well known~\cite{Opdam,Heckman} that there exists an intertwining operator~$M=M^{(g)}$ 
which establishes an isospectrality of $H^{(g)}$ and $H^{(g+1)}$. 
This differential operators of order $\sfrac12n(n{-}1)$  has  the simple form
\begin{equation} \label{M}
M \= \textrm{res}(\cM) \quad\with\quad
\cM \= \prod_{\mu<\nu} (\cD_\mu-\cD_\nu) \= \prod_{\alpha\in{\cal R}_+} \alpha\cdot\cD\ ,
\end{equation}
and using the Weyl antisymmetry of $\cM$ it is straightforward to verify that
(see e.g.~\cite{Feigin03})
\begin{equation} \label{MH}
[ \cM,\cH ] = 0 
\qquad\Longrightarrow\qquad
M^{(g)} H^{(g)} \= H^{(g+1)} M^{(g)}
\und
M^{(g)} \Psi_{\E,\ell}^{(g)}\; \sim\; \Psi_{\E,\ell}^{(g+1)}
\end{equation}
so that 
\begin{equation}
M^{(g)}\ : \quad \bigl(g\,,\,\ell\,,\,q\bigr)\ \mapsto\ 
\bigl(g{+}1\,,\,\ell\,,\,q{+}\sfrac12n(n{-}1)\bigr)
\end{equation}
simultaneously acts on the radial wave function $\sqrt{r}J_{q+(n-3)/2}$ and on the
angular wave function~$v_\ell^{(g)}$. Up to $n{=}4$ the expressions for $M^{(g)}$
were worked out explicitly in~\cite{CoLePl13}.

However, the angular system $H_\Om$ comes with its own intertwiners~$M_s=M_s^{(g)}$ of some order $s\in\NN$, 
which then provides {\it additional\/} intertwiners for~$H$.
This is suggested by the form (\ref{q}) of the quantum number~$q$ appearing in~(\ref{HOmspec}),
which reveals a partial\footnote{meaning that the intertwiners have a nontrivial kernel} 
isospectrality for $H_\Om$,
\begin{equation}
M_s^{(g)}\ : \quad \bigl(g\,,\,\ell\,,\,q\bigr)\ \mapsto\ 
\bigl(g{+}1\,,\,\ell{-}\sfrac12n(n{-}1)\,,\,q\bigr)\ .
\end{equation}
Indeed, as shown in~\cite{Feigin03}, one can more or less copy the strategy from $\R^{n-1}$ to $S^{n-2}$
and obtain the lowest-order angular intertwiner, with $s=\sfrac12(n{-}1)(n{-}2)$,
\begin{equation} \label{MOm}
\begin{aligned}
M_\Om &\ \equiv\ M_{\frac12(n-1)(n-2)} \= \textrm{res}(\cM_\Om) \qquad\with \\[4pt]
\cM_\Om &\= \sum_{\pi\in{\cal S}_n} \textrm{sgn}(\pi)\ \pi\Bigl(
\prod_{2\le\mu<\nu\le n}\bigl\{(x^1{-}x^\mu)(\cD_1{-}\cD_\nu)-
(x^1{-}x^\nu)(\cD_1{-}\cD_\mu)\bigr\}\Bigr) \\[4pt]
&\= \sum_{\pi\in{\cal S}_n} \textrm{sgn}(\pi)\ \pi\Bigl(
\prod_{2\le\mu<\nu\le n}\bigl\{\cL_{\mu 1}+\cL_{1\nu}-\cL_{\mu\nu}\bigr\}\Bigr)\ ,
\end{aligned}
\end{equation}
where the choice of $x^1$ as a reference is arbitrary and irrelevant.
Because $\cH$ is in the center of the algebra generated by $\{\cL_{ij}\}$
and $\cM_\Om$ is again antisymmetric under Weyl reflections,
one argues, analogously to (\ref{MH}) and with the help of $H^{(-g)}=H^{(g+1)}$, 
that~\cite{Feigin03}
\begin{equation} \label{MOH}
[ \cL_{ij},\cH ] = 0
\qquad\Longrightarrow\qquad
M_\Om^{(g)} H^{(g)} \= H^{(g+1)} M_\Om^{(g)}
\und 
M_\Om^{(g)} \Psi_{\E,\ell}^{(g)}\; \sim\; \Psi_{\E,\ell-n(n-1)/2}^{(g+1)}\ .
\end{equation}
These properties hold not only for $M_\Om$, but actually
any Weyl-{\it antiinvariant\/} function of $\{\cL_{ij}\}$ provides an angular intertwiner $M_s^{(g)}$
which obeys (\ref{MOH}).
By the same token, any Weyl-{\it invariant\/} function of $\{\cL_{ij}\}$ will yield, via its residue,
a conserved angular quantity~$C_t$ of some order~$t$.
We remark that the angular Dunkl operators do not commute with each other but 
form a subalgebra of a rational Cherednik algebra~\cite{FeHa14}.
It is a deformation of the $so(n{-}1)$ Lie algebra generated by $\{L_{ij}\}$.
The general problem of identifying all Liouville charges for the angular system reduces to
constructing a maximal abelian subalgebra in the algebra of Weyl-symmetric polynomials in~$\{\cL_{ij}\}$,
which is not an easy task. 
Likewise, identifying the minimal independent set of Weyl-antiinvariants and hence intertwiners is nontrivial.

Now, since any such $M_s$ and $C_t$ is scale invariant, i.e.\ $r$~independent,
it does not touch the radial wave function, and hence it is also true that
\begin{equation}
\begin{aligned}
{}[\cL_{ij},\cH_\Om] = 0
\qquad\Longrightarrow\qquad 
M_s^{(g)} H_\Om^{(g)} &\= H_\Om^{(g+1)} M_s^{(g)}
\and
M_s^{(g)} v_{\ell}^{(g)}\ \sim\ v_{\ell-n(n-1)/2}^{(g+1)} \\[4pt]
\textrm{as well as} \qquad 
C_t^{(g)} H_\Om^{(g)} &\= H_\Om^{(g)} C_t^{(g)}\ .
\end{aligned}
\end{equation}
To summarize, we have the connection
\begin{equation}
\begin{aligned}
{\cal C}_t(\cL_{ij}) \quad \textrm{Weyl-invariant} & \qquad\Longrightarrow\qquad 
C_t = \textrm{res}({\cal C}_t) \quad \textrm{commutes with} \ H_\Om \ ,\\[4pt]
\cM_s(\cL_{ij}) \quad \textrm{Weyl-antiinvariant} & \qquad\Longrightarrow\qquad 
M_s = \textrm{res}(\cM_s) \quad \textrm{intertwines with} \ H_\Om \ .
\end{aligned}
\end{equation}
In general, the angular conserved charges $C_t$ will not be in involution with one another,
but some combinations may be.
The angular intertwiners $M_s$ are differential operators of order~$s$ 
and have, in contrast to $M$, a sizeable kernel, with
\begin{equation}
\textrm{dim ker}_n(\ell) \= \textrm{deg}_n(\ell)\ -\ \textrm{deg}_n(\ell-\sfrac12n(n{-}1))\ .
\end{equation}
For small values of~$n$, we can be explicit:
\begin{equation} \label{dimker}
\begin{aligned}
\textrm{dim ker}_3(\ell) &\= \begin{cases} 1 &\for \ell=0 \\ 0 &\for \ell>0 \end{cases} \ ,\\
\textrm{dim ker}_4(\ell) &\= \begin{cases} 
1 &\for \ell=0,3 \ \textrm{mod} \ 4 \\ 0 &\for \ell=1,2 \ \textrm{mod} \ 4 \end{cases} \ ,\\
\textrm{dim ker}_5(\ell) &\= \Bigl\lfloor\frac{\ell}{6}\Bigr\rfloor \ +\ \begin{cases} 
1 &\for \ell=0,3,4,5,8,9,11 \ \textrm{mod} \ 12 \\
0 &\for \ell=1,2,6,7,10 \ \textrm{mod} \ 12 \end{cases} \ .
\end{aligned}
\end{equation}

With the help of the adjoint intertwiners, which reverse the direction of $M$ resp.~$M_s$,
\begin{equation}
{M_s^{(g)}}^\+ = M_s^{(-g)}
\qquad\Longrightarrow\qquad
M_s^{(-g)} H_\Om^{(g+1)} = H_\Om^{(g)} M_s^{(-g)}
\ \and\
M_s^{(-g)} v_{\ell}^{(g+1)} \sim v_{\ell+n(n-1)/2}^{(g)}\ ,
\end{equation}
it is trivial to construct special conserved charges,
\begin{equation}
R_{2s}^{(g)}\ :=\ M_s^{(-g)} M_s^{(g)}
\und
R_{2s}^{(1-g)} \= M_s^{(g-1)} M_s^{(1-g)}\ , 
\end{equation}
which provide particular cases of some $C_t$ for $t{=}2s$.
In the full Calogero model, $R=M^\+ M$ turned out to be a specific polynomial in the Liouville charges,
which was evaluated by infinitely separating the particles thus making contact with the free case.
This move is not possible for $R_{2s}$ due to the compactness of~$S^{n-2}$. 
Neither is it true that $R_{2s}^{(-g)}=R_{2s}^{(g+1)}$, and so these charges intertwine as follows,
\begin{equation}
M_s^{(g)} R_{2s}^{(g)} \= R_{2s}^{(-g)} M_s^{(g)} 
\= \bigl( R_{2s}^{(g+1)} - (1{+}2g)\rho_{2s}^{(g+1)} \bigr)\, M_s^{(g)}\ ,
\end{equation}
and similarly in the opposite direction. 
Here, $\rho_{2s}$ is a particular expression in other conserved charges.

Quite generally, we cannot expect conserved charges other than the angular Hamiltonian 
to be invariant under $g\to1{-}g$, and in fact one finds that
\begin{equation}
C_t^{(-g)} \ \neq\ C_t^{(g+1)} \und
M_s^{(g)} C_t^{(g)} \ \neq\ C_t^{(g+1)} M_s^{(g)}\ .
\end{equation}
Therefore, higher conserved quantities $C_t$ intertwine in a more complicated fashion,
\begin{equation} \label{generaltwine}
M_s^{(g)}\,C_t^{(g)} \= \sum_{s',t'} \Gamma_{st}^{s't'}(g)\,C_{t'}^{(g+1)} \,M_{s'}^{(g)}\ ,
\end{equation}
where $\Gamma_{st}^{s't'}(g)$ are certain polynomials in $g$ of order $s{+}t{-}s'{-}t'{-}1$ at the most.
The sum runs over all linearly independent conserved quantities including 
\begin{equation}
C_2^{(g)} \= -\textrm{res}\bigl(\cL^2\bigr)  \und C_0^{(g)} \ :=\ 1\ .
\end{equation}

When the coupling $g$ is an integer, repreated intertwining relates all quantities with their
analogs in the free theory ($g{=}0$ or $1$), which allows one to generate analytic expressions
for all wave functions. Moreover, in this case additional conserved charges make the model
`analytically integrable' and produce a `supercomplete' graded ring of commuting differential
operators~\cite{ChaVes90,Chalykh96,CoLePl13}.

\section{Warmup: the hexagonal or P\"oschl-Teller model}

\noindent
Let us work out and test the general assertions of the previous section on the simplest
nontrivial case, which occurs for $n=3$. The three-particle (or $A_2$) rational Calogero model
is long known to be completely separable, the ensuing angular (relative) model being the
famed P\"oschl-Teller model of a particle on a circle with a $\cos^{-2}(3\phi)$ potential.

We start by introducing Jacobi coordinates $(y^1,y^2)$ for the relative motion
(and the center of mass~$X$),
\begin{equation}
\begin{aligned}
x^1&\=X + \sfrac1{\sqrt{2}}\,y^1 + \sfrac1{\sqrt{6}}\,y^2  \quad,\qquad &
\pa_{x^1}&\=\sfrac13\pa_X+ \sfrac1{\sqrt{2}}\,\pa_{y^1} + \sfrac1{\sqrt{6}}\,\pa_{y^2} \quad, \\[4pt]
x^2&\=X - \sfrac1{\sqrt{2}}\,y^1 + \sfrac1{\sqrt{6}}\,y^2  \quad,\qquad &
\pa_{x^2}&\=\sfrac13\pa_X- \sfrac1{\sqrt{2}}\,\pa_{y^1} + \sfrac1{\sqrt{6}}\,\pa_{y^2}\quad, \\[4pt]
x^3&\=X - \sfrac2{\sqrt{6}}\,y^2 \quad,\qquad &
\pa_{x^3}&\=\sfrac13\pa_X  - \sfrac2{\sqrt{6}}\,\pa_{y^2} \quad, 
\end{aligned}
\end{equation}

and then polar $(r,\phi)$ and complex $(w,\wb)$ coordinates on $\R^2$,
\begin{equation}
y^1 \= r\,\cos\phi \and y^2 \= r\,\sin\phi 
\qquad\Longrightarrow\qquad w\ :=\ y^1+\im y^2 \= r\,\ep^{\im\phi}\ .
\end{equation}
After the standard similarity transformation, the (reduced) Hamiltonian 
\begin{equation}
\begin{aligned}
H &\= -\sfrac12(\pa_1^2+\pa_2^2)\ +\ g(g{-}1)\Bigl(
\frac2{(2\,y^1)^2}+\frac2{(y^1{-}\sqrt{3}y^2)^2}+\frac2{(y^1{+}\sqrt{3}y^2)^2}\Bigr) \\[4pt]
&\= -\sfrac12(\pa_1^2+\pa_2^2)\ +\ 
\frac{9\,g(g{-}1)\,((y^1)^2{+}(y^2)^2)}{2\,(y^1)^2\,(y^1{-}\sqrt{3}y^2)^2\,(y^1{+}\sqrt{3}y^2)^2}
\end{aligned}
\end{equation}
and its angular cousin
take the following form as differential operators on $\R^2$ and $S^1$, respectively,
\begin{equation}
H'\,u\ \mapsto\ -\half\bigl(\pa_r^2+\sfrac1{4\,r^2})\,u\ +\ \sfrac1{r^2} H_\Om\,u
\qquad\textrm{with}\qquad H_\Om \= -\half\pa_\phi^2 + U\ ,
\end{equation}
where the angular potential receives contributions from the three positive roots of~$A_2$,
\begin{equation}
U(\phi) \= 
\sfrac{g(g-1)}2\bigl( \cos^{-2}(\phi)+\cos^{-2}(\phi{+}\sfrac{2\pi}3)+\cos^{-2}(\phi{-}\sfrac{2\pi}3) \bigr)
\= \sfrac92 g(g{-}1)\,\cos^{-2}(3\phi)\ .
\end{equation}
The poles separate the circle into six disjoint arcs, 
pertaining to the hexagonal symmetry of the $A_2$ root system.
We display the angular Hamiltonian in complex coordinates as well,
\begin{equation}
H_\Om \= \half\bigl(w\pa_w-\wb\pa_\wb\bigr)^2\ +\ g(g{-}1)\frac{18\,(w\wb)^3}{(w^3+\wb^3)^2}\ .
\end{equation}

Specializing (\ref{HOmspec})--(\ref{deg}) to $n{=}3$, one finds that
the angular momentum~$\ell=3\ell_3$ is a multiple of three and
\begin{equation}
\ve_q \= \half q^2 \qquad\textrm{with}\qquad q \= 3g+\ell \= 3(g+\ell_3) \und \textrm{deg}(\ve_q)=1\ .
\end{equation}
The $\sqrt{r}$ factor in (\ref{radfct}) gets cancelled in the similarity transformation (\ref{sim}), giving
\begin{equation}
\Psi_{\E,q}(r,\phi) \= J_q({\scriptstyle\sqrt{2E}}\,r)\,v_q(\phi)\ ,
\end{equation}
and the angular wave function reads
\begin{equation} \label{vellgPT}
v_q(\phi) \ \equiv\ v_\ell^{(g)}(\phi) \ \sim\ 
r^{q}\,\Bigl(\sum_{\nu=1}^3 \cD_\nu^3\Bigr)^{\ell_3}\,\De^g\,r^{-6g} \ \sim\
r^q\,\bigl(\cD_w^3-\cD_\wb^3\bigr)^{\ell_3}\,\De^g\,r^{-6g}
\end{equation}
with
\begin{equation}
\De\ \sim\ (x^1{-}x^2)(x^2{-}x^3)(x^3{-}x^1)\ \sim\ r^3\,\cos(3\phi) \ \sim\ w^3+\wb^3\ ,
\end{equation}
where we have rewritten in complex coordinates.\footnote{
We refrain from displaying $v_q(\phi)$ in Jacobi or polar coordinates as the expressions are not enlightening.}
Owing to the $S_3$ Weyl group of~$A_2$, we take profit from the complex cubic roots of unity
to cast the holomorphic Dunkl operator $\cD_w=\sfrac12(\cD_1-\im\cD_2)$ into the simple form
\begin{equation}
\cD_w \= \pa_w\ -\ g\,\Bigl\{ 
\frac{1}{w+\wb}\,s_0 + \frac{\rho}{\rho w+\rb\wb}\,s_+ + \frac{\rb}{\rb w+\rho\wb}\,s_- \Bigr\}
\with \rho=\ep^{2\pi\im/3}\ ,
\end{equation}
where the three basic Weyl reflections act as follows in the complex plane,
\begin{equation}
s_0 : \ w\ \mapsto\ -\wb \quad,\qquad
s_+ : \ w\ \mapsto\ -\rho\wb \quad,\qquad
s_- : \ w\ \mapsto\ -\rb\wb\quad,
\end{equation}
or on the polar angle,
\begin{equation}
s_0 : \ \phi\ \mapsto\ \pi-\phi \quad,\qquad
s_+ : \ \phi\ \mapsto\ -\sfrac{\pi}{3}-\phi \quad,\qquad
s_- : \ \phi\ \mapsto\ \sfrac{\pi}{3}-\phi \quad.
\end{equation}

According to (\ref{M}), the full intertwiner is a third-order operator,\footnote{
We remark that $\cD_w^3+\cD_\wb^3$ changes sign under any Weyl reflection while 
$\cD_w^3-\cD_\wb^3$ is totally symmetric.}
\begin{equation}
\cM\ \sim\ (\cD_1-\cD_2)(\cD_2-\cD_3)(\cD_3-\cD_1) \ \sim\
\cD_w^3 + \cD_\wb^3\ ,
\end{equation}
whose evaluation eventually produces~\cite{CoLePl13}
\begin{equation}
\begin{aligned}
M&\ \sim\ \pa_{12}\pa_{23}\pa_{31}
-\sfrac{2g}{x^{12}}\pa_{23}\pa_{31}
-\sfrac{2g}{x^{23}}\pa_{31}\pa_{12}
-\sfrac{2g}{x^{31}}\pa_{12}\pa_{23} \\[4pt] & \ \
+\sfrac{4g^2}{x^{12}x^{23}}\pa_{31}
+\sfrac{4g^2}{x^{23}x^{31}}\pa_{12}
+\sfrac{4g^2}{x^{31}x^{12}}\pa_{23}
-\sfrac{g(g{-}1)}{(x^{12})^2}\pa_{12}
-\sfrac{g(g{-}1)}{(x^{23})^2}\pa_{23}
-\sfrac{g(g{-}1)}{(x^{31})^2}\pa_{31} \\[4pt] & \ \
-\sfrac{6\,g^2(g{+}1)}{x^{12}x^{23}x^{31}}
+2g(g{-}1)(g{+}2)\Bigl(\sfrac{1}{(x^{12})^3}+
\sfrac{1}{(x^{23})^3}+\sfrac{1}{(x^{31})^3}\Bigr)\ ,
\end{aligned}
\end{equation}
with the abbreviated notation $x^{\mu\nu}\equiv x^\mu{-}x^\nu$ and
$\pa_{\mu\nu}\equiv\pa_{x^\mu}{-}\pa_{x^\nu}$.
In complex form, it reads
\begin{equation}
\begin{aligned}
M\ &\sim\ \pa_w^3+\pa_\wb^3 - \frac{6\,g}{w^3{+}\wb^3}\bigl(
w^2\pa^2_w + w\wb\,\pa_w\pa_\wb + \wb^2\pa^2_\wb\bigr) +
\frac{6g(3g{-}1)}{w^3{+}\wb^3}\bigl(w\pa_w+\wb\pa_\wb\bigr) \\[4pt]
&\ -\ \frac{9g(g{-}1)}{(w^3{+}\wb^3)^2}\bigl(w^4\pa_w+\wb^4\pa_\wb\bigr) +
\frac{3g(9g{+}14)}{w^3{+}\wb^3} - 
\frac{27g(g{+}2)}{(w^3{+}\wb^3)^3}\bigl(w^6{+}\wb^6+2g\,w^3\wb^3\bigr)\ ,
\end{aligned}
\end{equation}
and in the polar representation the intertwiner becomes
\begin{equation}
\begin{aligned}
M \ &\sim \cos3\phi\,\Bigl\{ 
    \pa_r^3 - \sfrac{3(3g+1)}{r}\pa_r^2 + \sfrac{3(9g+1)}{r^2}\pa_r -
    \sfrac{24g}{r^3} + \sfrac{6}{r^2}H_\Om\pa_r - \sfrac{6(g+2)}{r^3}H_\Om\Bigr\} \\
&\ -\ \sin3\phi\,M_\Om\,\Bigl\{ 
    \sfrac{3}{r}\pa_r^2 - \sfrac{9}{r^2}\pa_r + \sfrac{8}{r^3} + \sfrac{2}{r^3}H_\Om\Bigr\}\ ,
\end{aligned}
\end{equation}
where the angular intertwiner~$M_\Om$ (to be given shortly) has been used.

Reducing to the angular subsystem, i.e.~the P\"oschl-Teller model, 
the formula~(\ref{MOm}) yields
\begin{equation}
\cM_\Om \equiv \cM_1 \ \sim\ x^1\cD_2-x^2\cD_1+x^2\cD_3-x^3\cD_2+x^3\cD_1-x^1\cD_3 
\ \sim\ \im \bigl( w\cD_w-\wb\cD_\wb \bigr)\ ,
\end{equation}
which simplifies to
\begin{equation}
\cM_1\ \sim\ x^1\pa_{23}+x^2\pa_{31}+x^3\pa_{12}\ +\ g\,\Bigl\{ 
\frac{x^1{+}x^2{-}2x^3}{x^{12}}s_{12} +
\frac{x^2{+}x^3{-}2x^1}{x^{23}}s_{23} +
\frac{x^3{+}x^1{-}2x^2}{x^{31}}s_{31} \Bigr\}
\end{equation}
or to
\begin{equation}
\cM_1\ \sim\ \im \bigl( w\pa_w-\wb\pa_\wb \bigr)\ -\ \im g\,\Bigl\{
\frac{w-\wb}{w+\wb}\,s_0 +\frac{\rho w-\rb\wb}{\rho w+\rb\wb}\,s_+ 
+\frac{\rb w-\rho\wb}{\rb w+\rho\wb}\,s_-\Bigr\}\ .
\end{equation}
Taking the residue, we arrive at
\begin{equation}
M_\Om \equiv M_1\  \sim\ 
x^1\pa_{23}+x^2\pa_{31}+x^3\pa_{12}\ -\ g\,
\frac{(x^1{+}x^2{-}2x^3)(x^2{+}x^3{-}2x^1)(x^3{+}x^1{-}2x^2)}{x^{12}\,x^{23}\,x^{31}}
\end{equation}
or, in complex and polar parametrizations, 
\begin{equation}
M_1\ \sim\ 
\im \bigl( w\pa_w-\wb\pa_\wb \bigr)\ -\ 3\im g\,\frac{w^3-\wb^3}{w^3+\wb^3} \=
\pa_\phi + 3g\,\tan3\phi \= \cos^g(3\phi)\,\pa_\phi\,\cos^{-g}(3\phi)\ .
\end{equation}
The penultimate expression is the familiar first-order operator
which enjoys the following remarkable properties:
\begin{itemize}
\addtolength{\itemsep}{-6pt}
\item it adds zeros at $\phi=0,\sfrac{2\pi}3,\sfrac{4\pi}{3}$ and poles at $\phi=\sfrac{\pi}{3},\pi,\sfrac{5\pi}{3}$
\item it intertwines between symmetric and antisymmetric Weyl group singlets
\item acting $\ell_3{+}1$ times on $v_\ell^{(0)}$ gives zero
\item acting with $M_1^{(1)}=\smash{{M_1^{(-1)}}^\+}$ on $v_\ell^{(0)}$ 
      yields non-normalizable wave functions $v_\ell^{(-1)}$
\item it is a Vandermonde dressing of the angular momentum
\item it is the unique angular intertwiner
\end{itemize}

For $g{=}0$, the expression (\ref{vellgPT}) is readily evaluated 
(with $r^0\to\ln r$ and $\ell=3\ell_3$),
\begin{equation}
v_\ell^{(0)} \ \sim\ r^\ell\,(\pa_w^3-\pa_\wb^3)^{\ell_3}\,\ln w\wb \ \sim\
r^\ell\,(w^{-\ell} \pm \wb^{-\ell}) \ \sim\ \begin{cases} 
\cos\ell\phi & \textrm{for $\ell$ even} \\[4pt]
\sin\ell\phi & \textrm{for $\ell$ odd} \end{cases} \ ,
\end{equation}
which are all Weyl symmetric.
Acting repeatedly with $M_1$ on these wave functions, one alternates between 
Weyl symmetric and Weyl antisymmetric and
generates the wave functions at the same value of~$\ell$ for any integer coupling.
The low-lying states for $g=0,1,2,3$ are tabulated in Table~1.
The only exception is the one-dimensional kernel of $M_1$, consisting of the ground state
\begin{equation}
v_0^{(g)} \ \sim\ r^{-3g}\,\De^g \ \sim\ \cos^g 3\phi\ .
\end{equation}
\begin{figure}[h!]
\centering
\includegraphics[scale=0.92]{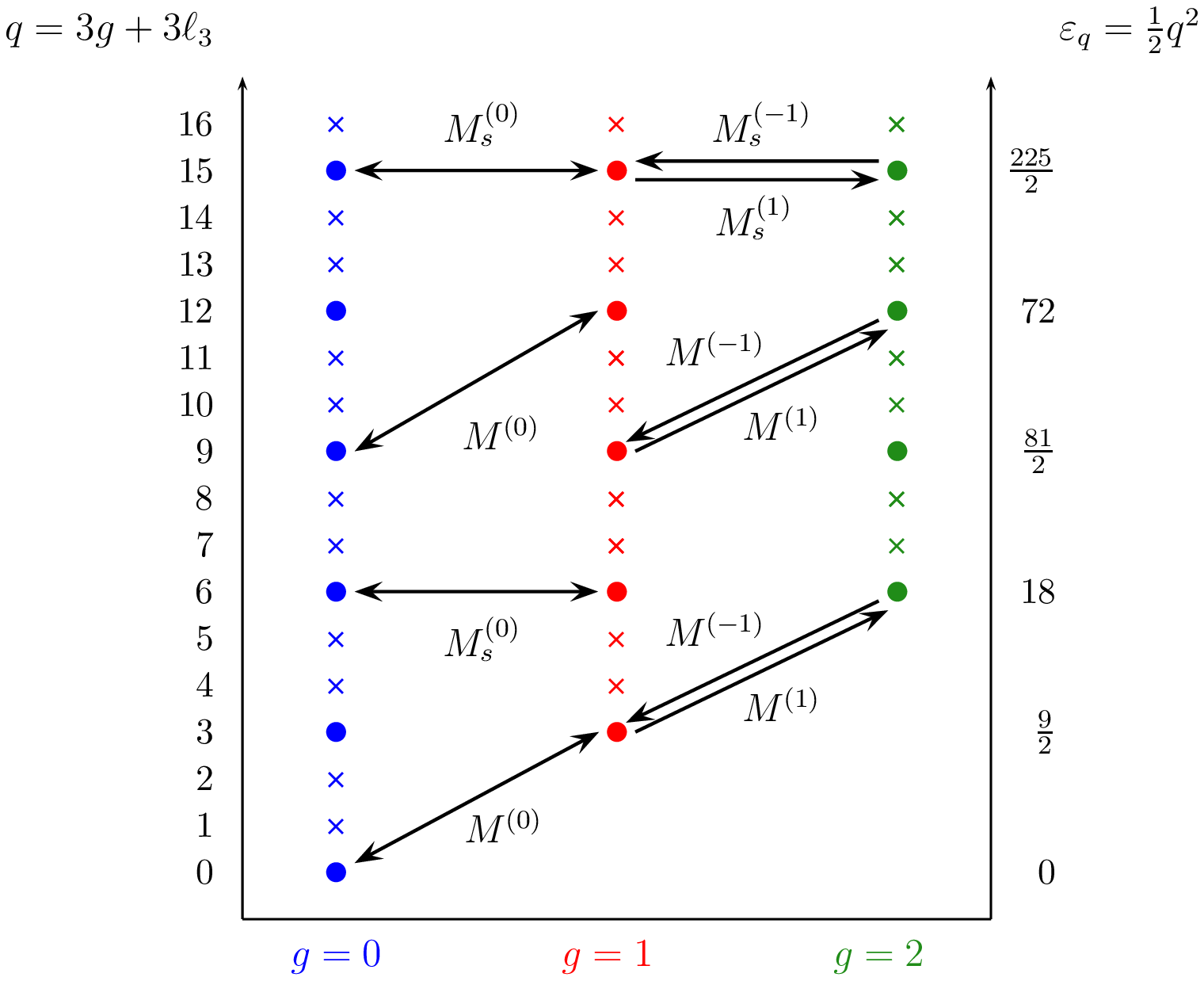}
\caption{P\"oschl-Teller model: spectrum and intertwiner actions.}
\label{plots1}
\end{figure}
Finally, the additional conserved quantity~$R_2$ manufactured with $M_\Om$ and its adjoint,
yields nothing new since
\begin{equation}
R_2^{(g)}\= (\pa_\phi - 3g\,\tan3\phi)(\pa_\phi + 3g\,\tan3\phi) \= -2\,H_\Om^{(g)} + 9\,g^2
\= -\textrm{res}(\cL^2) \= -C_2^{(g)}\ ,
\end{equation}
which provides $\rho_2^{(g+1)}=9$.
We read off the intertwining relation
\begin{equation}
M_1^{(g)}\,R_2^{(g)}\= \bigl( R_2^{(g+1)} - 9(1{+}2g)\bigr)\,M_1^{(g)} \ .
\end{equation}

For generic values of~$g$, there can be no further independent conserved quantity.
However, when $g$ is integer, one additional algebraically independent conserved charge 
can be constructed,
\begin{equation}
Q^{(g)} \= M_1^{(g-1)}M_1^{(g-2)}\cdots M_1^{(1-g)}
\qquad\Longrightarrow\qquad
Q^{(g)} H_\Om^{(g)} \= Q^{(g)} H_\Om^{(1-g)} \= H_\Om^{(g)} Q^{(g)}\ ,
\end{equation}
which is a Weyl-antiinvariant differential operator of order $2g{-}1$ that squares to
\begin{equation}
\bigl(Q^{(g)}\bigr)^2 \= \smash{\prod_{j=1-g}^{g-1}}\bigl(-2\,H_\Om^{(g)}+9j^2\bigr)\ .
\end{equation}

\newpage

\section{Tetrahexahedric model: the spectrum}

\noindent
The main purpose of this paper is to work out the angular (relative) rational
4-particle (or $A_3$) Calogero model as the simplest non-separable case,
in order to demonstrate the viability of the general observations of Section~2.
We take advantage of the isometry $A_3\simeq D_3$ and employ $D_3$-type relative 
coordinates.\footnote{
See also \cite{Dunkl08} for an application of the same idea.}

\begin{figure}[h!]
\centering
\includegraphics[scale=0.20]{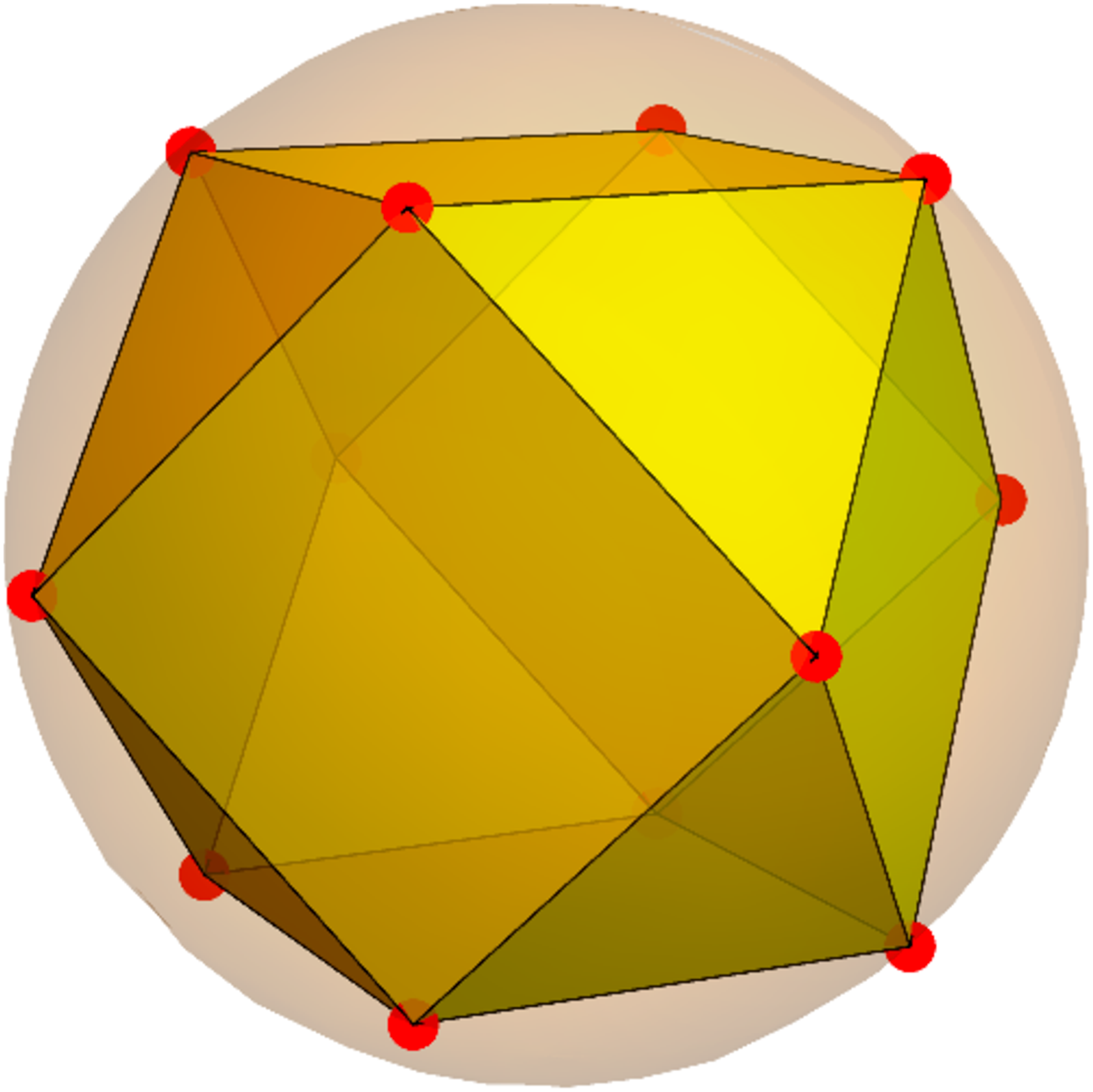}
\includegraphics[scale=0.20]{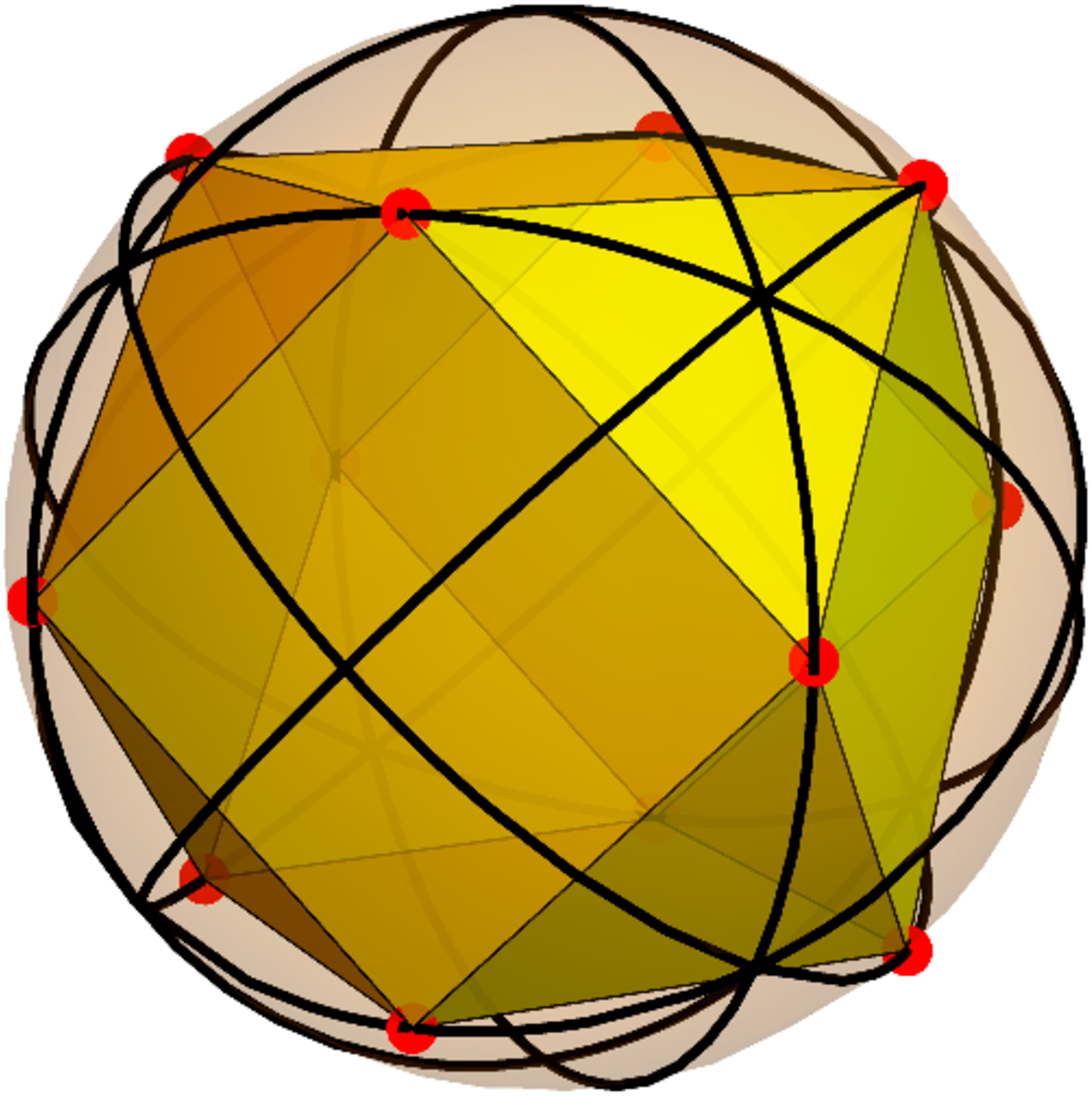}
\includegraphics[scale=0.20]{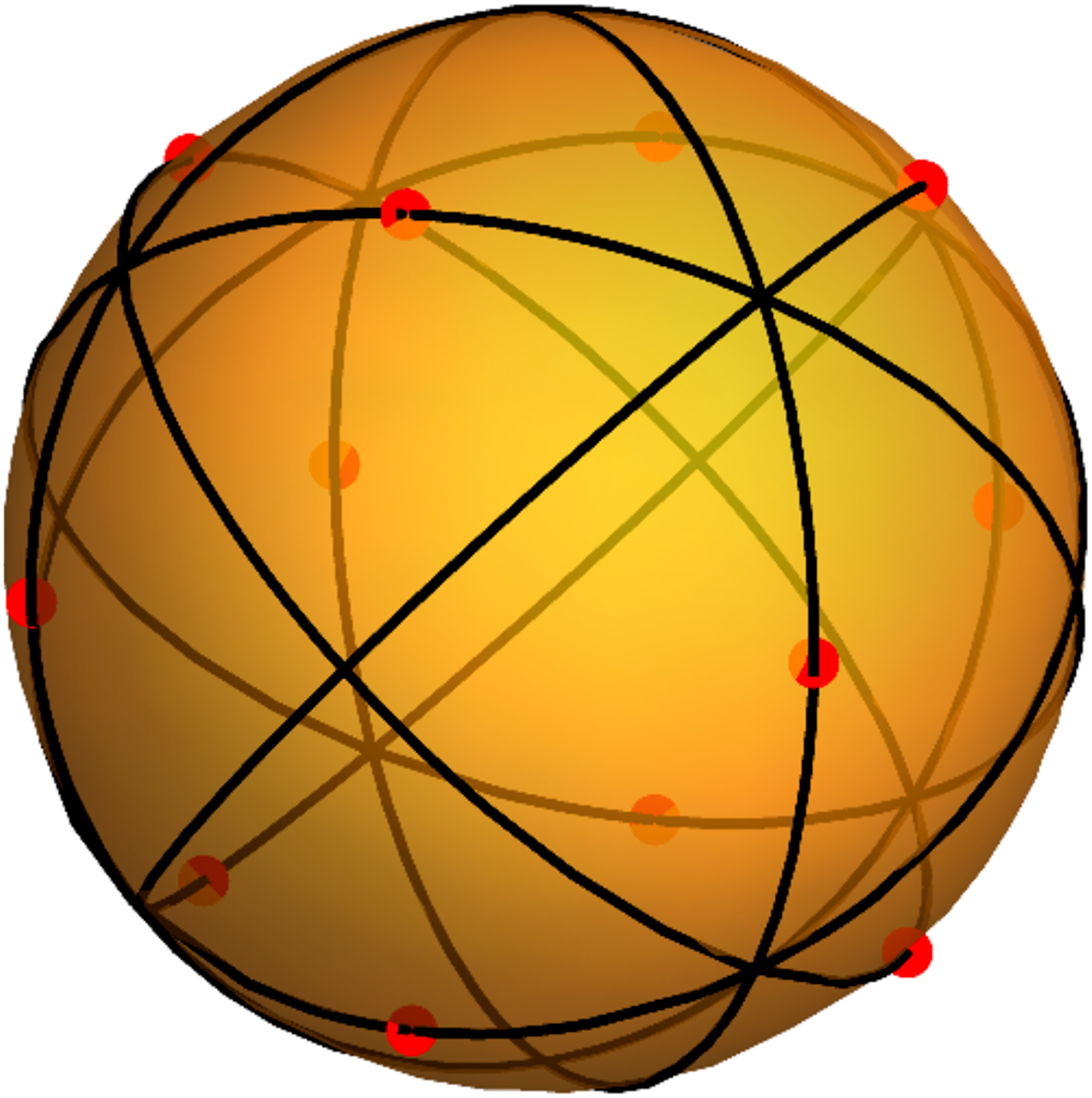}
\includegraphics[scale=0.20]{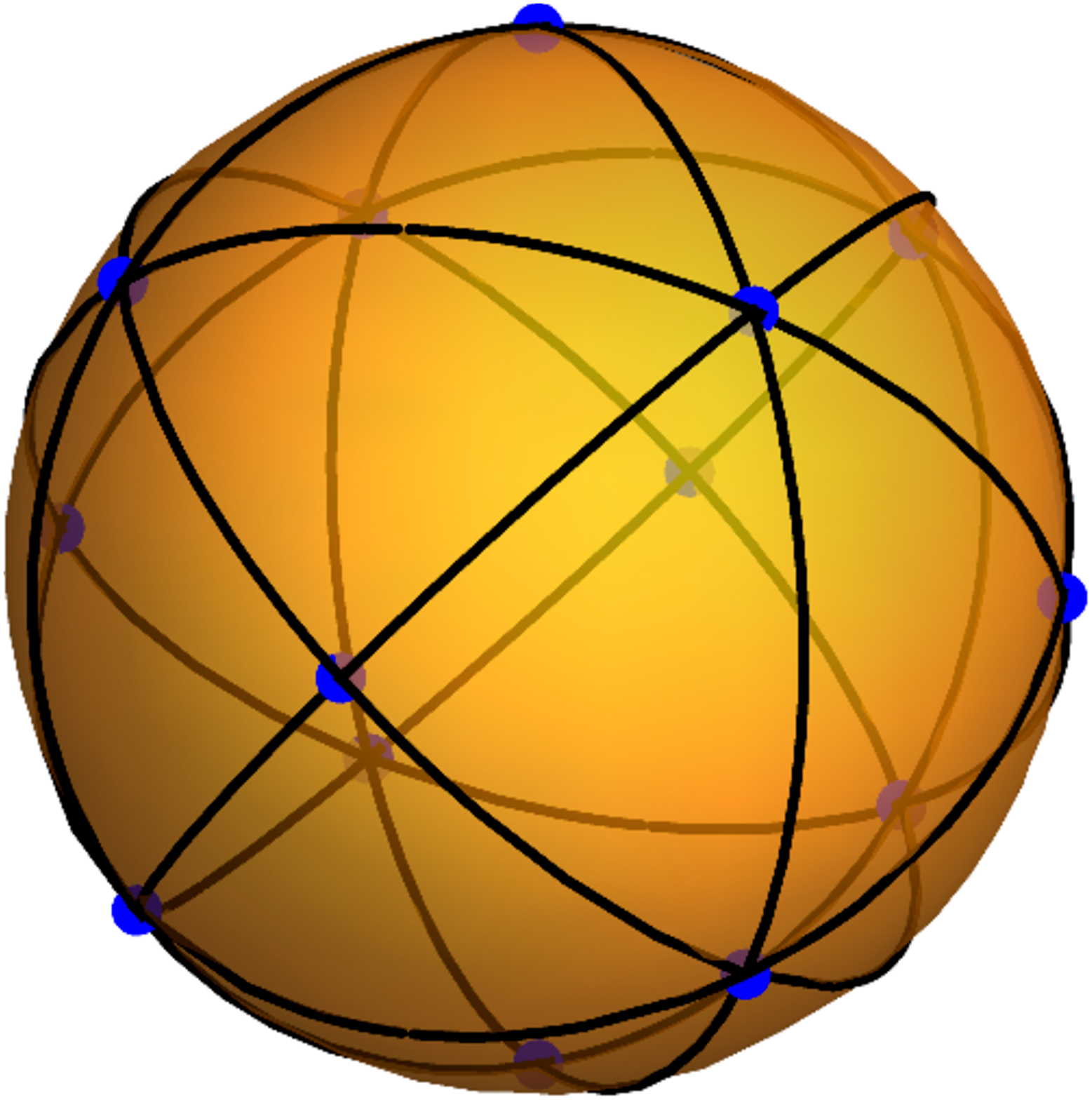}
\includegraphics[scale=0.20]{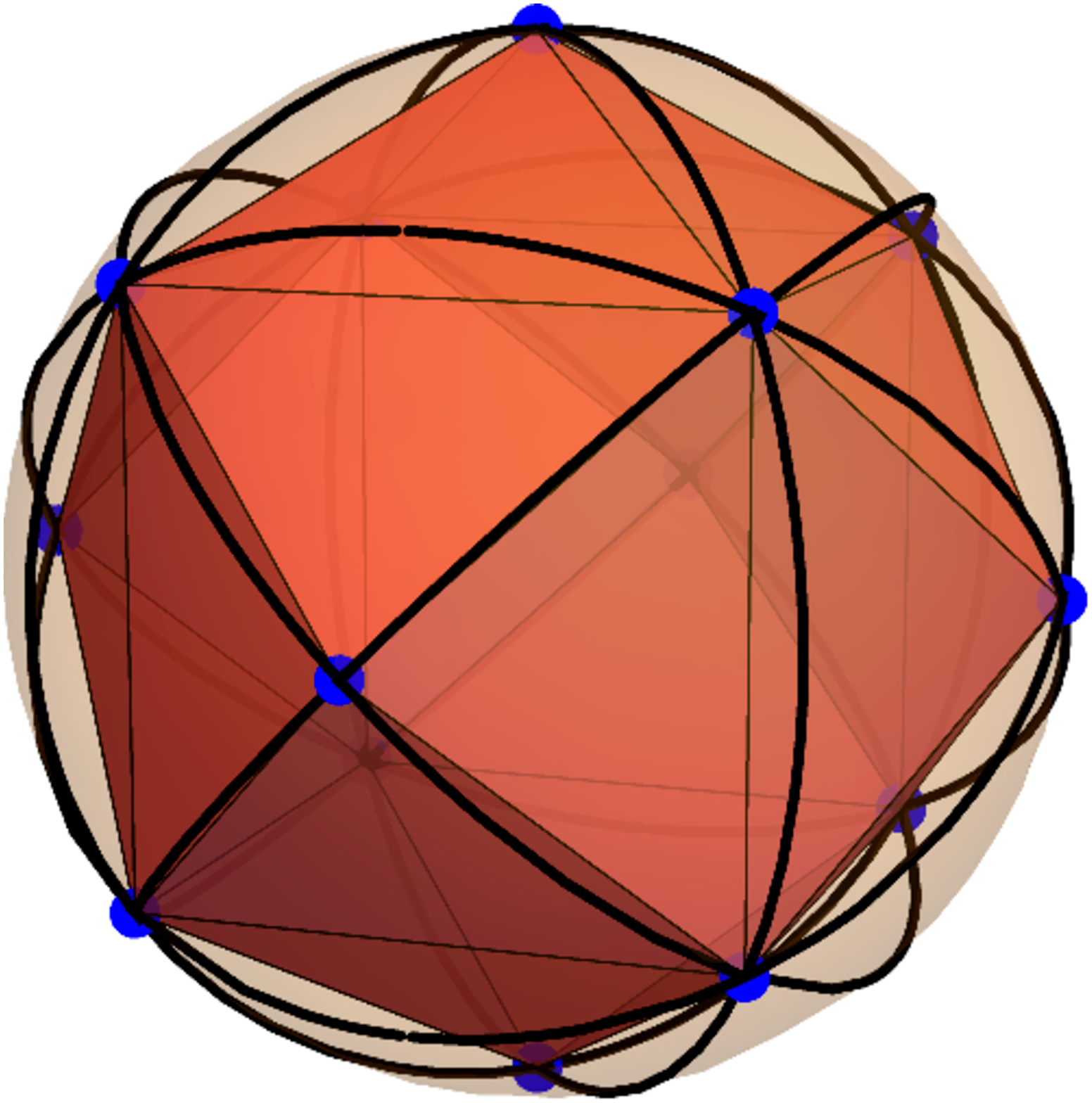}
\includegraphics[scale=0.20]{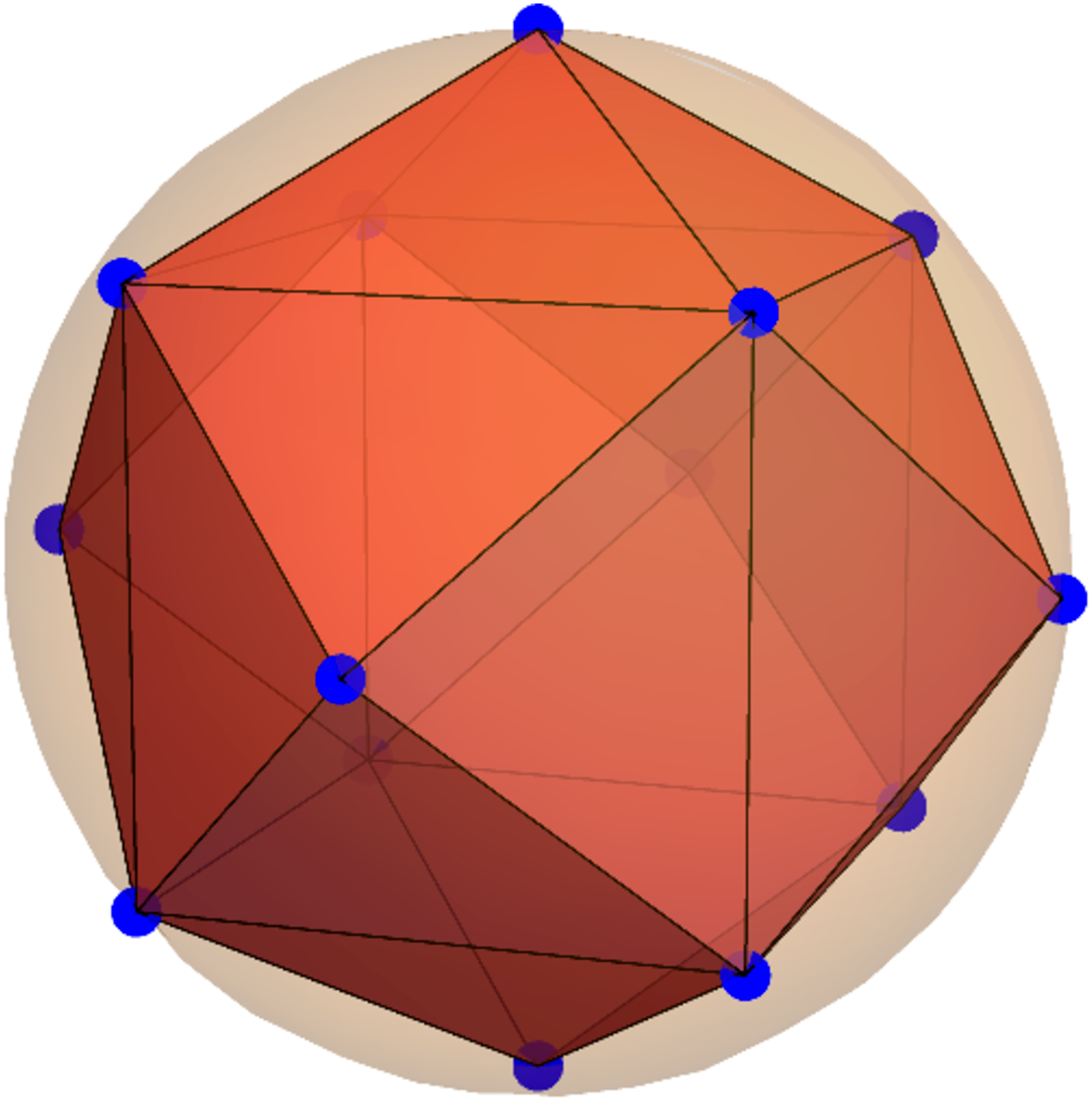}
\caption{From the cuboctahedron to the tetrahexahedron.}
\label{polyhedra}
\end{figure}

\begin{figure}[h!]
\centering
\includegraphics[scale=0.2]{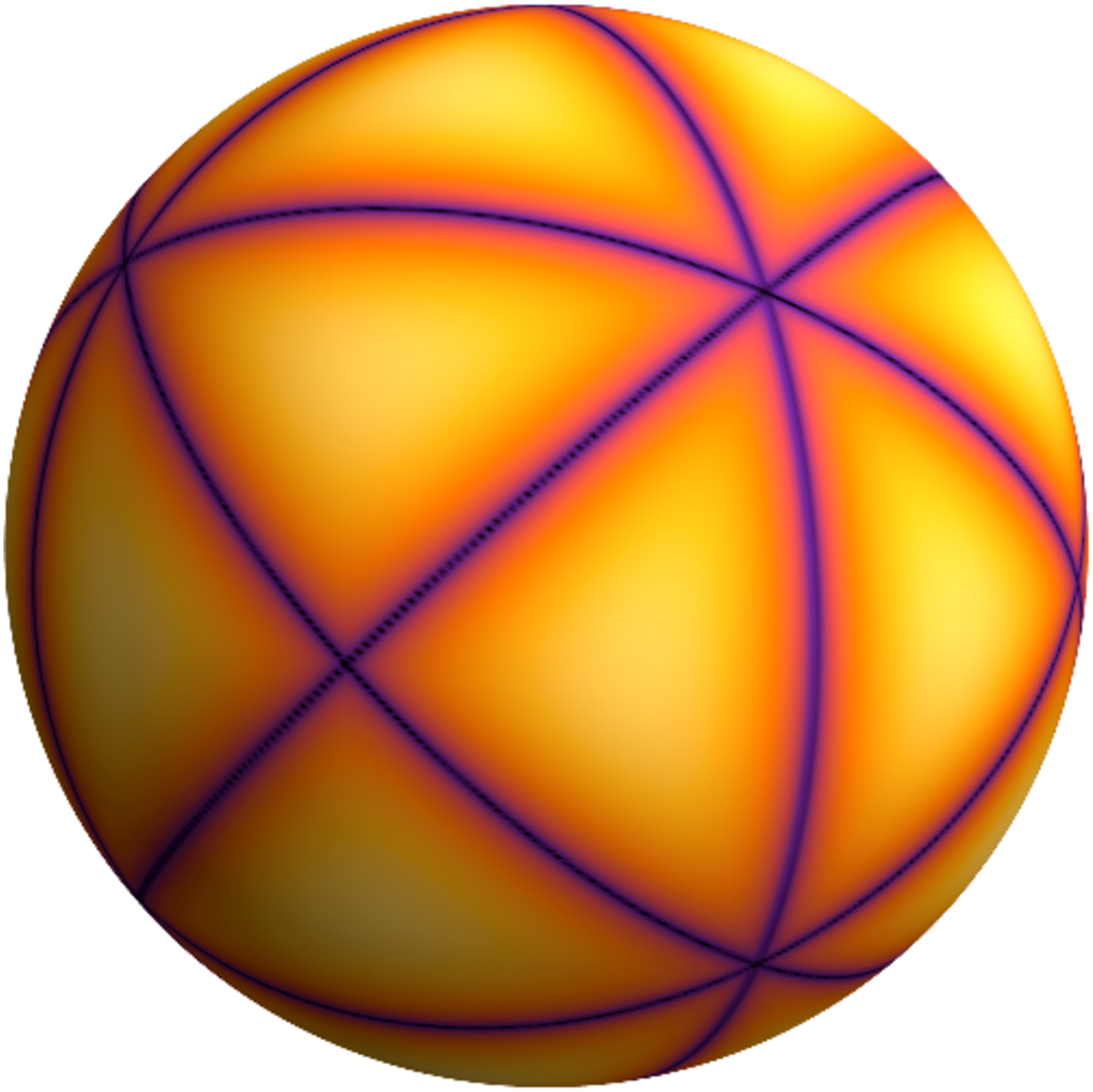}\quad\quad\quad
\includegraphics[scale=0.20]{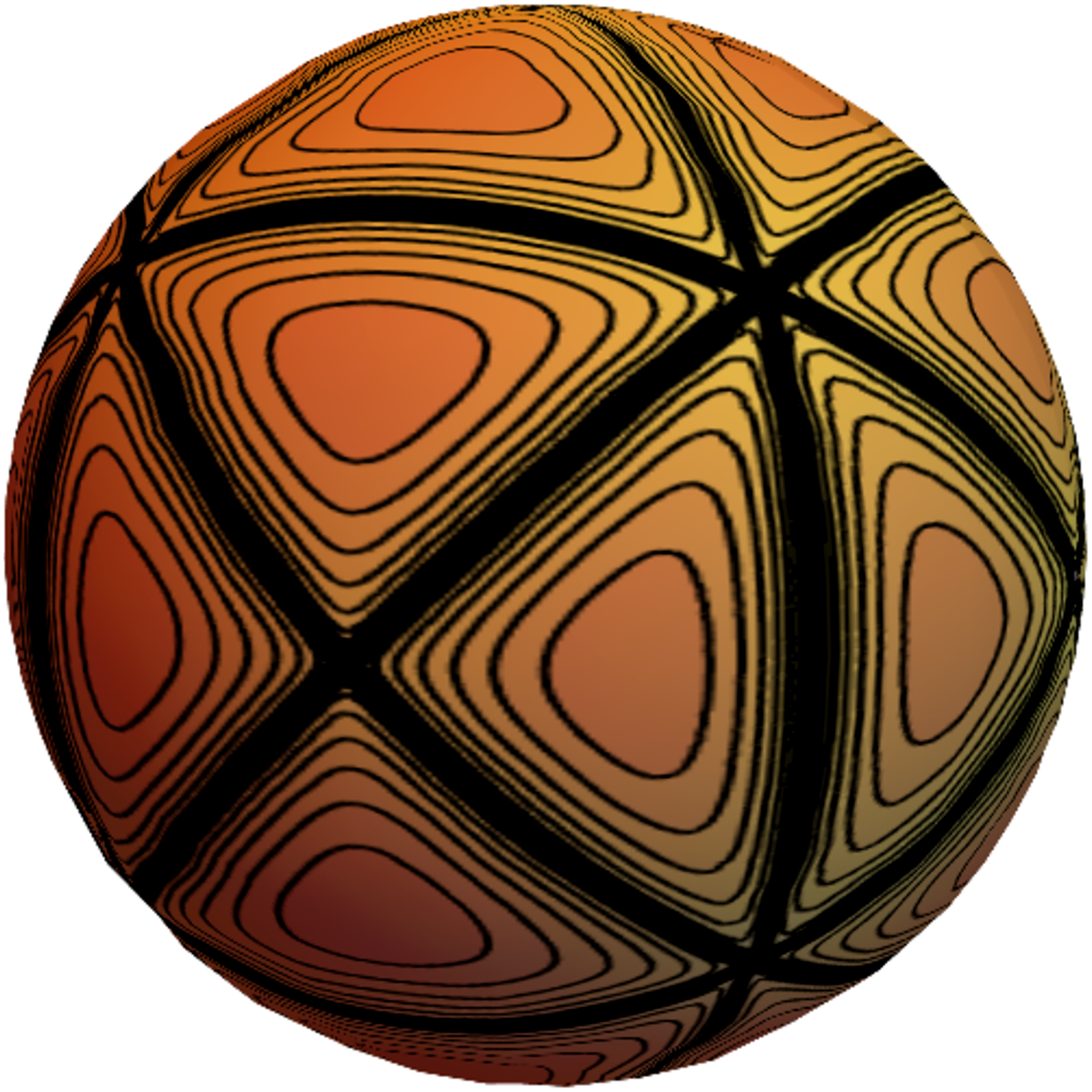}
\caption{The tetrahexahedric potential (its log log log) on the two-sphere.}
\label{potentialplot}
\end{figure}

To construct the model and find its spectrum, we introduce not Jacobi but
Walsh-Hadamard coordinates $\{y^i\}$ and their derivatives together with the center of mass~$X$ via
\begin{equation}
\begin{aligned}
x^1\=X+\half(+y^1+y^2+y^3) \quad,\qquad &
\pa_{x^1}\=\sfrac14\pa_X+\half(+\pa_{y^1}+\pa_{y^2}+\pa_{y^3}) \quad, \\[4pt]
x^2\=X+\half(+y^1-y^2-y^3) \quad,\qquad &
\pa_{x^2}\=\sfrac14\pa_X+\half(+\pa_{y^1}-\pa_{y^2}-\pa_{y^3}) \quad, \\[4pt]
x^3\=X+\half(-y^1+y^2-y^3) \quad,\qquad &
\pa_{x^3}\=\sfrac14\pa_X+\half(-\pa_{y^1}+\pa_{y^2}-\pa_{y^3}) \quad, \\[4pt]
x^4\=X+\half(-y^1-y^2+y^3) \quad,\qquad &
\pa_{x^4}\=\sfrac14\pa_X+\half(-\pa_{y^1}-\pa_{y^2}+\pa_{y^3}) \quad.
\end{aligned}
\end{equation}
The polar (or spherical) parametrization of $\R^3$ defines the radius~$r$ and two angles $\th$ and $\phi$,
\begin{equation}
y^1 \= r\,\sin\th\,\cos\phi \ ,\quad
y^2 \= r\,\sin\th\,\sin\phi \ ,\quad
y^3 \= r\,\cos\th \ .
\end{equation}
To avoid index cluttering, we redenote the Euclidean $\R^3$ coordinates in a traditional fashion,
\begin{equation}
(y^1\,,\,y^2\,,\,y^3) \ =:\ (x\,,\,y\,,\,z)\ .
\end{equation}
The angular momentum components read (up to a factor of~$\im$)
\begin{equation} \label{LTH}
L_x = -(y\pa_z{-}z\pa_y) \ ,\qquad
L_y = -(z\pa_x{-}x\pa_z) \ ,\qquad
L_z = -(x\pa_y{-}y\pa_x) \ ,
\end{equation}
so that (minus) the Laplacian on $S^2$ takes the form
\begin{equation}
L^2 \= -(L_x^2+L_y^2+L_z^2) \=
-\sfrac{1}{\sin\th}\pa_\th\sin\th\,\pa_\th - \sfrac{1}{\sin^2\th}\pa_\phi^2\ ,
\end{equation}
and the (reduced) Hamiltonian becomes
\begin{equation}
H \= -\half(\pa_x^2+\pa_y^2+\pa_z^2)\ +\ 2\,g(g{-}1)\Bigl(
\frac{x^2+y^2}{(x^2-y^2)^2} + \frac{y^2+z^2}{(y^2-z^2)^2} + \frac{z^2+x^2}{(z^2-x^2)^2}\Bigr)\ .
\end{equation}
After the standard similarity transformation, we obtain
\begin{equation}
H'\,u\ \mapsto\ -\half\pa_r^2\,u\ +\ \sfrac1{r^2} H_\Om\,u
\und  H_\Om \= \half L^2 + U
\end{equation}
with 
\begin{equation}
U(\th,\phi) \= 2g(g{-}1)\biggl\{
\frac1{\sin^2\th\,\cos^22\phi}+
\frac{\cos^2\th+\sin^2\th\cos^2\phi}{(\cos^2\th-\sin^2\th\cos^2\phi)^2}+
\frac{\cos^2\th+\sin^2\th\sin^2\phi}{(\cos^2\th-\sin^2\th\sin^2\phi)^2} \biggr\}\ .
\end{equation}
This has been named `cuboctahedric Higgs oscillator' potential~\cite{HNY} since it is
the superposition of six Higgs oscillators placed at the positive roots of~$A_3\simeq D_3$,
and those (together with their antipodes) form the vertices of a cuboctahedron~\cite{cuboctahedron}.
However, a better name is `tetrahexahedral' potential, because the intersections of the
Weyl chamber walls with the two-sphere form six great circles, which trace the 36 edges
of a tetrahexahedron (or tetrakis hexahedron)~\cite{tetrahexahedron}. 
Its 24 identical faces are right isosceles triangles. The spherical projection tessellates
the two-sphere into 24 spherically triangular domains in which the particle is trapped.
This sequence of geometrical configurations, starting from the $A_3$ root system (cuboctahedron)
and ending at the walls of the angular potential (tetrahexahedron) is displayed in Figure~\ref{polyhedra},
and color and contour plots of the potential are shown in Figure~\ref{potentialplot}.

The angular energy eigenvalues become
\begin{equation}
\ve_q \= \half q\,(q{+}1) \qquad\textrm{with}\qquad q \= 6g+\ell \= 6g+3\ell_3{+}4\ell_4\ .
\end{equation}
The interaction reduces the $2\ell{+}1$ dimensional free eigenspace 
(spanned by the spherical harmonics $Y_{\ell m}$) to the subspace invariant under the
$S_4$ Weyl group action generated e.g.~by
\begin{equation}
\begin{aligned}
& s_{x+y} : \ (x,y,z)\mapsto(-y,-x,+z) \quad,\qquad
  s_{x-y} : \ (x,y,z)\mapsto(+y,+x,+z) \quad,\\[4pt]
& s_{y+z} : \ (x,y,z)\mapsto(+x,-z,-y) \quad,\qquad
  s_{y-z} : \ (x,y,z)\mapsto(+x,+z,+y) \quad,\\[4pt]
& s_{z+x} : \ (x,y,z)\mapsto(-z,+y,-x) \quad,\qquad 
  s_{z-x} : \ (x,y,z)\mapsto(+z,+y,+x) \quad.
\end{aligned}
\end{equation}
This reduces the degeneracy to $\textrm{deg}(\ve_q)=\textrm{deg}_4(\ell)$ of (\ref{degsmall}).
The corresponding wave functions
\begin{equation}
\Psi_{\E,q}(r,\th,\phi) \= j_q({\scriptstyle\sqrt{2E}}\,r)\,v_q(\th,\phi)
\end{equation}
have a spherical Bessel function as their radial part and involve two specific Weyl-symmetric
homogeneous Dunkl polynomials, of degree three and four,
\begin{equation} \label{vellgTH}
\begin{aligned}
v_q(\th,\phi)\ \equiv\ v_\ell^{(g)}(\th,\phi) &\ \sim\
r^{q+1}\,\Bigl(\sum_{\mu=1}^3 \cD_\mu^3\Bigr)^{\ell_3}\,
\Bigl(\sum_{\nu=1}^3 \cD_\nu^4\Bigr)^{\ell_4}\, \De^g\,r^{-1-12g}  \\[4pt]
&\ \sim\ r^{q+1}\,\bigl(\cD_x\cD_y\cD_z\bigr)^{\ell_3}\,
\bigl(\cD_x^4{+}\cD_y^4{+}\cD_z^4\bigr)^{\ell_4}\,\De^g\,r^{1-12g}\ ,
\end{aligned}
\end{equation}
as well as the coordinate Vandermonde,
\begin{equation}
\begin{aligned}
\De &\= \prod_{i<j}\bigl((y^i)^2-(y^j)^2\bigr) \= (x^2-y^2)(x^2-z^2)(y^2-z^2) \\
&\ \sim\ r^6\,\sin^2\th\,\cos^4\th\,\cos^22\phi\,(\tan^2\th\cos^2\phi-1)(\tan^2\th\sin^2\phi-1)\ .
\end{aligned}
\end{equation}
Here, the three Dunkl operators on $\R^3$ are given by
\begin{equation} \label{DunklTH}
\begin{aligned}
\cD_x&\=\pa_x\ -\ \frac{g}{x{+}y}\,s_{x+y}-\frac{g}{x{-}y}\,s_{x-y}-
\frac{g}{x{+}z}\,s_{x+z}-\frac{g}{x{-}z}\,s_{z-x} \ ,\\[4pt]
\cD_y&\=\pa_y\ -\ \frac{g}{y{+}x}\,s_{x+y}-\frac{g}{y{-}x}\,s_{x-y}-
\frac{g}{y{+}z}\,s_{y+z}-\frac{g}{y{-}z}\,s_{y-z} \ ,\\[4pt]
\cD_z&\=\pa_z\ -\ \frac{g}{z{+}x}\,s_{z+x}-\frac{g}{z{-}x}\,s_{z-x}-
\frac{g}{z{+}y}\,s_{y+z}-\frac{g}{z{-}y}\,s_{y-z} \ .
\end{aligned}
\end{equation}
The angular wave functions $v_q(\th,\phi)$ in (\ref{vellgTH}) are known as
`tetrahedral harmonics' in theoretical chemistry~\cite{FoxOzier70,Wormer01}.\footnote{
The chemistry literature seems to be unaware 
of the straightforward construction scheme~(\ref{vellgTH}).}
The low-lying states for $g=0,1,2$ are tabulated in Table~2.

For completeness, we here display some quantities in the potential-free frame.
The potential-free Hamiltonian
\begin{equation}
\widetilde{H} \= -\half(\pa_x^2+\pa_y^2+\pa_z^2)\ -\ 
\bigl( \sfrac{2g\,x}{x^2{-}y^2}+\sfrac{2g\,x}{x^2{-}z^2} \bigr)\pa_x -
\bigl( \sfrac{2g\,y}{y^2{-}z^2}+\sfrac{2g\,y}{y^2{-}x^2} \bigr)\pa_y -
\bigl( \sfrac{2g\,z}{z^2{-}x^2}+\sfrac{2g\,z}{z^2{-}y^2} \bigr)\pa_z 
\end{equation}
annihilates the corresponding Dunkl-deformed harmonics
\begin{equation}
\widetilde{h}_{\ell_3,\ell_4}^{(g)}\ \sim\
r^{12g+1+6\ell_3+8\ell_4}\,\Delta^{-g}\,\bigl(\cD_x\cD_y\cD_z\bigr)^{\ell_3}\,
\bigl(\cD_x^4{+}\cD_y^4{+}\cD_z^4\bigr)^{\ell_4}\,\De^g\,r^{1-12g}\ ,
\end{equation}
so that the complete wave function may be written as
\begin{equation}
\Psi^{(g)}_{\E,\ell_3,\ell_4}(x,y,z) \= j_q({\scriptstyle\sqrt{2E}}\,r)\,r^{-q}\De^g\,
\widetilde{h}^{(g)}_{\ell_3,\ell_4}(x,y,z) 
\qquad\textrm{with}\qquad q=6g+3\ell_3+4\ell_4\ .
\end{equation}

\section{Tetrahexahedric model: intertwiners \&\ integrability}

According to (\ref{M}), the four-particle intertwiner is
\begin{equation}
M \ \sim\ \textrm{res}\bigl((\cD_x^2-\cD_y^2)(\cD_y^2-\cD_z^2)(\cD_z^2-\cD_x^2)\bigr)\ ,
\end{equation}
and it acts on the wave functions as
\begin{equation}
M^{(g)}\,j_q({\scriptstyle\sqrt{2E}}\,r)\,v_\ell^{(g)}(\th,\phi)\ \sim\ 
j_{q+6}({\scriptstyle\sqrt{2E}}\,r)\,v_\ell^{(g+1)}(\th,\phi)
\end{equation}
The rather complicated explicit form of this six-order partial differential operator
has been constructed in~\cite{CoLePl13}.
Let us turn our attention to the angular intertwiners~$M_s$. Their building blocks
are the Dunkl-deformed angular momenta
\begin{equation}
\cL_x \equiv \cL_{yz} = -(y\cD_z{-}z\cD_y) \ ,\quad
\cL_y \equiv \cL_{zx} = -(z\cD_x{-}x\cD_z) \ ,\quad
\cL_z \equiv \cL_{xy} = -(x\cD_y{-}y\cD_x)
\end{equation}
or their complex combinations $\cL_\pm=\cL_x\pm\im\cL_y$.
Employing (\ref{LTH}) and (\ref{DunklTH}), they read
\begin{equation}
\begin{aligned}
\cL_x &\= L_x\ +\ g\bigl\{
\sfrac{z}{x-y}s_{x-y}-\sfrac{z}{x+y}s_{x+y}-\sfrac{y+z}{y-z}s_{y-z}+
\sfrac{y-z}{y+z}s_{y+z}+\sfrac{y}{z-x}s_{z-x}+\sfrac{y}{z+x}s_{z+x} \bigr\} \ ,\\[4pt]
\cL_y &\= L_y\ +\ g\bigl\{
\sfrac{x}{y-z}s_{y-z}-\sfrac{x}{y+z}s_{y+z}-\sfrac{z+x}{z-x}s_{z-x}+
\sfrac{z-x}{z+x}s_{z+x}+\sfrac{z}{x-y}s_{x-y}+\sfrac{z}{y+x}s_{x+y} \bigr\} \ ,\\[4pt]
\cL_z &\= L_z\ +\ g\bigl\{
\sfrac{y}{z-x}s_{z-x}-\sfrac{y}{z+x}s_{z+x}-\sfrac{x+y}{x-y}s_{x-y}+
\sfrac{x-y}{x+y}s_{x+y}+\sfrac{x}{y-z}s_{y-z}+\sfrac{x}{z+y}s_{y+z} \bigr\}
\end{aligned}
\end{equation}
and transform under the Weyl reflections in the following fashion,
\begin{equation} \label{WeylcL}
\begin{aligned}
& s_{x+y} : \ (\cL_x,\cL_y,\cL_z)\mapsto(+\cL_y,+\cL_x,-\cL_z) \ ,\quad
  s_{x-y} : \ (\cL_x,\cL_y,\cL_z)\mapsto(-\cL_y,-\cL_x,-\cL_z) \ ,\\[4pt]
& s_{y+z} : \ (\cL_x,\cL_y,\cL_z)\mapsto(-\cL_x,+\cL_z,+\cL_y) \ ,\quad
  s_{y-z} : \ (\cL_x,\cL_y,\cL_z)\mapsto(-\cL_x,-\cL_z,-\cL_y) \ ,\\[4pt]
& s_{z+x} : \ (\cL_x,\cL_y,\cL_z)\mapsto(+\cL_z,-\cL_y,+\cL_x) \ ,\quad 
  s_{z-x} : \ (\cL_x,\cL_y,\cL_z)\mapsto(-\cL_z,-\cL_y,-\cL_x) \ .
\end{aligned}
\end{equation}
For convenience, we also note their transformation under certain even Weyl group elements,
\begin{equation} \label{WeylcLeven}
\begin{aligned}
& s_{x+y}s_{x-y} : \ (\cL_x,\cL_y,\cL_z)\mapsto(-\cL_x,-\cL_y,+\cL_z) \ ,\\[4pt]
& s_{y+z}s_{y-z} : \ (\cL_x,\cL_y,\cL_z)\mapsto(+\cL_x,-\cL_y,-\cL_z) \ ,\\[4pt]
& s_{z+x}s_{z-x} : \ (\cL_x,\cL_y,\cL_z)\mapsto(-\cL_x,+\cL_y,-\cL_z) \ .\quad 
\end{aligned}
\end{equation}

The construction (\ref{MOm}) then yields a simple cubic polynomial in the
angular Dunkl operators,
\begin{equation}
\cM_\Om\ \equiv\ \cM_3 \= \sfrac16\bigl(
\cL_x\cL_y\cL_z + \cL_x\cL_z\cL_y + \cL_y\cL_z\cL_x + 
\cL_y\cL_x\cL_z + \cL_z\cL_x\cL_y + \cL_z\cL_y\cL_x \bigr)\ ,
\end{equation}
which indeed is an anti-invariant under the Weyl group.
It remains to evaluate the residue $M_\Om\=\res\bigl(\cM_\Om\bigr)$ for arriving at
\begin{equation}
\begin{aligned}
M_3 &\= 
\sfrac16\bigl(L_xL_yL_z + L_xL_zL_y ) -\ 2g\sfrac{xy}{x^2{-}y^2}\,\bigl(L_xL_y+L_yL_x\bigr) \\[4pt]
&\ \ + \left[ 16 g^2\sfrac{ x^2}{\left(x^2-y^2\right) \left(z^2-x^2\right)}
-4 g(g{-}1) \sfrac{ x^2}{\left(y^2-z^2\right)^2}
+2 g \bigl(\sfrac{1}{x^2-y^2}-\sfrac{1}{z^2-x^2}\bigr) \right] y z\, L_x \\[4pt]
&\ \ +2 g(g{-}1)(g{+}2)\,x^2 \left[ \sfrac{ y^2+z^2}{(y^2-z^2)^2}
+ z \bigl(\sfrac{1}{(y-z)^3}-\sfrac{1}{(y+z)^3}\bigr)  \right] 
+g\left(2 g^2{+}8 g{-}1\right)\sfrac{ y^2+z^2}{y^2-z^2} \\[4pt]
&\ \ +2g^2(8{+}9g)\sfrac{ x^2 y^2 z^2}{(x^2 - y^2) (x^2 - z^2) (y^2 - z^2)}
-\sfrac{2}{3}g^3 \sfrac{ x^6+y^6+z^6}{\left(x^2-y^2\right) \left(x^2-z^2\right) \left(y^2-z^2\right)}
+\ \textrm{cyclic permutations}\ .
\end{aligned}
\end{equation}
Inserting the free angular momentum differential operators~(\ref{LTH}), we finally obtain
\begin{equation}
\begin{aligned}
M_3 &\= y^2z\pa_{zxx}-yz^2\pa_{xxy}+\half(y^2{-}z^2)\pa_{xx}+\ 
4g\,\sfrac{yz}{y^2{-}z^2}\bigl(yz\pa_{xx}-zx\pa_{xy}+x^2\pa_{yz}-xy\pa_{zx})  \\[4pt]
&\ \ +g \left[ 2\,y^2 z^2 \bigl(\sfrac{8 g}{\left(x^2-y^2\right) \left(z^2-x^2\right)}+
\sfrac{16 g}{\left(z^2-x^2\right) \left(y^2-z^2\right)}-\sfrac{2 g-1}{\left(x^2-y^2\right)^2}+
\sfrac{2 g-1}{\left(z^2-x^2\right)^2}\bigr)\right.  \\[4pt]
&\ \left. \qquad -\sfrac{2 x^2 y^2}{\left(z^2-x^2\right)^2}+\sfrac{2 x^2 z^2}{\left(x^2-y^2\right)^2}-
\sfrac{2 y^2}{x^2-y^2}-\sfrac{2 z^2}{z^2-x^2}-2\sfrac{ y^2+z^2}{y^2-z^2}\right] x \pa_{x} \\[4pt]
&\ \ +2 g(g{-}1)(g{+}2)\,x^2 \left[ \sfrac{ y^2+z^2}{\left(y^2-z^2\right)^2}+ z \bigl(\sfrac{1}{(y-z)^3}-
\sfrac{1}{(y+z)^3}\bigr)  \right] +g\left(2 g^2{+}8 g{-}1\right)\sfrac{ y^2+z^2}{y^2-z^2} \\[4pt]
&\ \ +2g^2(8{+}9g)\sfrac{ x^2 y^2 z^2}{(x^2 - y^2) (x^2 - z^2) (y^2 - z^2)}
-\sfrac{2}{3}g^3 \sfrac{ x^6+y^6+z^6}{\left(x^2-y^2\right) \left(x^2-z^2\right) \left(y^2-z^2\right)}
+\ \textrm{cyclic permutations}\ .
\end{aligned}
\end{equation}

This is not the only elementary intertwiner however. At order six, one finds the next 
Weyl-antiinvariant function of $\{\cL_x,\cL_y,\cL_z\}$, namely
\begin{equation}
\cM_6 \= 
\{\cL_x^4,\cL_y^2\}-\{\cL_y^4,\cL_x^2\}+\{\cL_y^4,\cL_z^2\}
-\{\cL_z^4,\cL_y^2\}+\{\cL_z^4,\cL_x^2\}-\{\cL_x^4,\cL_z^2\}\ .
\end{equation}
Its residue then produces a second intertwiner,
\begin{equation}
\begin{aligned}
M_6 &\= \left\{ L_x^4,L_y^2 \right\} + 
4g\bigl( \sfrac{x}{y+z}-\sfrac{x\,z}{z^2-x^2} \bigr) \left\{ L_x^4,L_y \right\} +
4g\bigl( \sfrac{y\,z}{x^2-y^2}+\sfrac{y(y-2z)}{y^2-z^2}\bigr) \left\{ L_x^3,L_y^2 \right\} \\[4pt]
&\ \ +\ O(L_x^4,L_x^3L_y,L_x^2L_y^2)\ \pm \text{permutations}\ .
\end{aligned}
\end{equation}
We have checked that all higher-order intertwiners, generated from further Weyl-antiinvariants,
can be decomposed into linear combinations of $M_3$ and $M_6$ with Weyl-symmetric functions of 
$\{\cL_x,\cL_y,\cL_z\}$ as coefficients.
In this sense, $M_3$ and $M_6$ algebraically generate the set of all intertwiners for $H_\Om$.
As a curiosity, we report a remarkable (non-local) example from this set,
\begin{equation}
\begin{aligned}
\cM_{\textrm{non-local}} &\= \sin\cL_+\sinh\cL_+ + \sin\cL_-\sinh\cL_- \=
\sum_{k=0}^\infty (-1)^k\frac{\cL_+^{4k+2}+\cL_-^{4k+2}}{(2k{+}1)!(4k{+}1)!!} \\[4pt]
&\= (\cL_{+}^2+\cL_-^2) - \sfrac1{90}(\cL_+^6{+}\cL_-^6) + 
\sfrac1{113400}(\cL_+^{10}{+}\cL_-^{10}) - \sfrac1{681080400}(\cL_+^{14}{+}\cL_-^{14}) +\ldots\ .
\end{aligned}
\end{equation}
In all cases, the action on the angular states is as expected,
\begin{equation}
M_s^{(g)}\,v_\ell^{(g)}(\th,\phi)\ \sim\ v_{\ell-6}^{(g+1)}(\th,\phi)\ ,
\end{equation}
and the size of the kernels $\textrm{dim ker}\,M_s=\textrm{dim ker}_4(\ell)$ is independent
of~$g$ as was given in~(\ref{dimker}).

\begin{figure}[h!]
\centering
\includegraphics[scale=0.46]{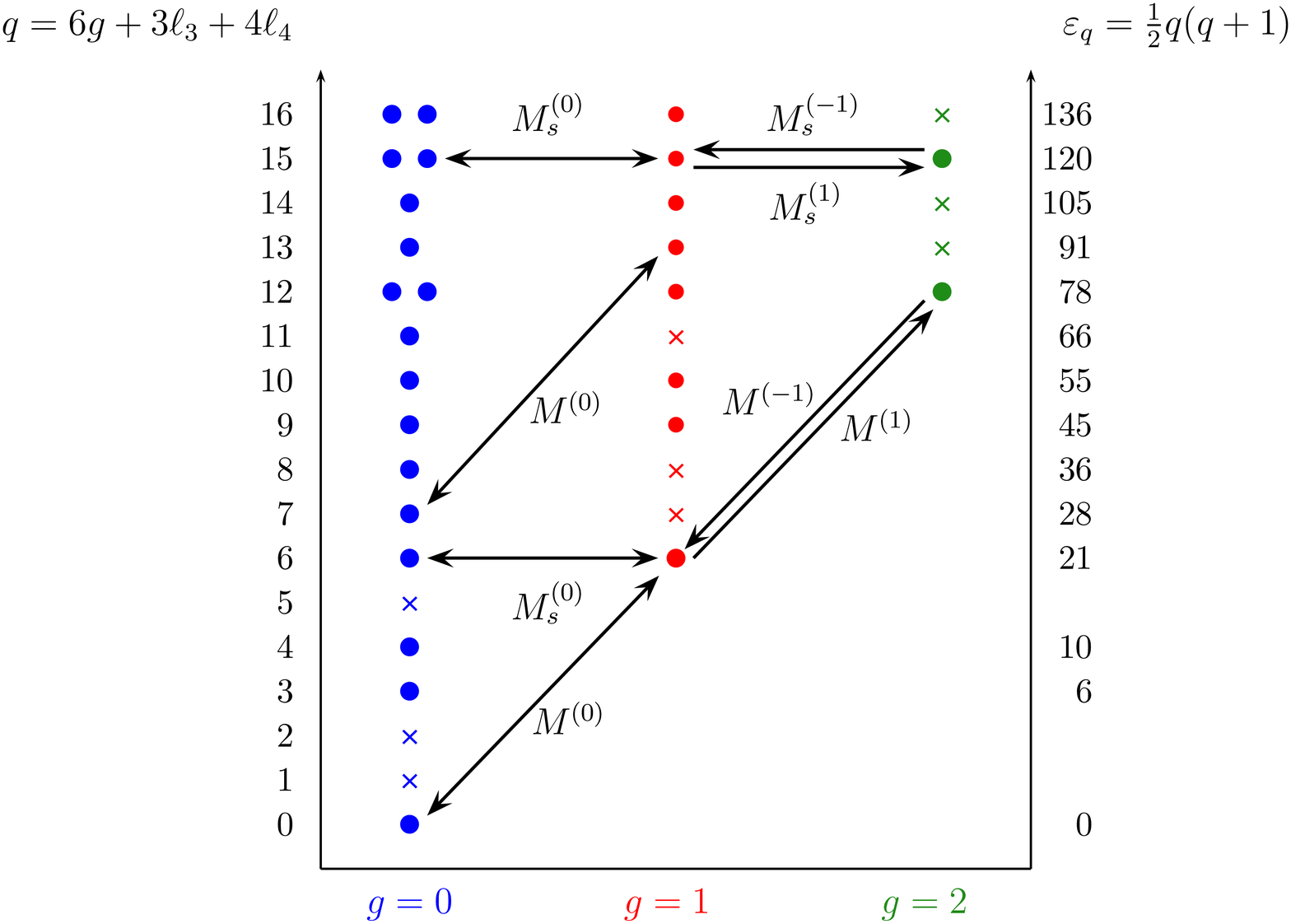}
\caption{Tetrahexahedric model: spectrum and intertwiner actions.}
\label{plots2}
\end{figure}

What about additional conserved charges in our tetrahexahedral angular Calogero system?
The $S_4$ Weyl group action~(\ref{WeylcL}) on the Dunkl angular momenta is sufficiently simple
to characterize the algebra of Weyl-symmetric polynomials. It is generated by
\begin{equation}
J_k \ :=\ \textrm{res}\bigl(\cL_x^k + \cL_y^k + \cL_z^k\bigr)
\quad\for k=0,2,4,6\ ,
\end{equation}
and its center is formed by
\begin{equation}
J_0 \= C_0 \= 1 \und
J_2 \= -C_2 \= -2\,H_\Om\ +\ 6g(6g{+}1)\ ,
\end{equation}
so $[J_4,H_\Om]=[J_6,H_\Om]=0$ but $[J_6,J_4]\neq0$.
Any word formed from $J_4$ and $J_6$ will commute with $H_\Om$.
The commutator $[J_6,J_4]$ cannot be expressed in terms of lower charges
but is related to a Weyl-invariant product of the two intertwiners,
\begin{equation}
\begin{aligned}
{}[J_6,J_4] 
\ &=\ 12 M_3^\+ M_6 + 24(3{+}4g)J_6J_2 -12(3{+}2g)J_4J_4 - 48(1{+}2g)J_4J_2J_2 \\
&\ + 12(1{+}2g)J_2J_2J_2J_2 + 24(97{+}113g{-}96g^2{-}144g^3)J_6 \\
&\ +12(266{+}349g{-}304g^2{-}528g^3)J_4J_2 + 12(76{+}101g{-}80g^2{-}144g^3)J_2J_2J_2 \\
&\ - 48(134{+}275g{-}1015g^2{-}1596g^3{+}1008g^4{+}1728g^5)J_4 \\
&\ + 24(146{+}379g{-}690g^2{-}1240g^3{+}672g^4{+}1152g^5)J_2J_2 \\
&\ + 96(10{-}39g{-}775g^2{-}2232g^3{-}1584g^4)J_2\ .
%\ \textrm{lower-order terms}\ ,
\end{aligned}
\end{equation}
Higher-degree invariants are not independent, e.g.\
\begin{equation}
\begin{aligned}
6J_8 \ &=\ 8J_6J_2+3J_4J_4-6J_4J_2J_2+J_2J_2J_2J_2 \\
&\ - 12(8{+}5g{+}12g^2) J_6 + 4(34{+}23g{+}30g^2) J_4J_2 - 8(5{+}3g{+}3g^2) J_2J_2J_2\\
&\ + 24(13{+}15g{-}102g^2{-}72g^3)J_4 - 4(43{+}70g{-}252g^2{-}144g^3) J_2J_2\\
&\ - 48(1{+}3g)(1{+}4g)(1{-}12g) J_2\ .
\end{aligned}
\end{equation}

It is instructive to work out the complete set of elementary intertwinings:
\begin{eqnarray}
M_3^{(g)} J_2^{(g)} &=& \bigl(J_2^{(g+1)}-6(7{+}12g)\bigr)\,M_3^{(g)} \ ,\\[4pt]
M_3^{(g)} J_4^{(g)} &=& \bigl(J_4^{(g+1)}-4(11{+}12g)J_2^{(g+1)}+48(26{+}73g{+}48g^2)\bigr)\,M_3^{(g)} 
\ +\ 2\,M_6^{(g)} \ ,
\quad\ \ {}
\end{eqnarray}
\begin{equation}
\begin{aligned}
M_3^{(g)} J_6^{(g)}\ &=\ \bigl(J_6^{(g+1)}-(35{+}36g)J_4^{(g+1)}-3(7{+}4g)J_2^{(g+1)}J_2^{(g+1)} \\
&\ +\ 2(1111{+}2668g{+}1392g^2)J_2^{(g+1)} \\
&\ +\ 96(457{+}1933g{+}2717g^2{+}1368g^3{+}144g^4)\bigr)M_3^{(g)} \\
&\ +\ \bigl(3J_2^{(g+1)}-(115{+}200g{+}48g^2)\bigr)M_6^{(g)}\ ,
\qquad\qquad\qquad\qquad\qquad\qquad\!\!{}
\end{aligned}
\end{equation}
\begin{equation}
M_6^{(g)} J_2^{(g)}\ =\ \bigl(J_2^{(g+1)}-6(7{+}12g)\bigr)\,M_6^{(g)} \ ,
\qquad\qquad\qquad\qquad\qquad\qquad\qquad\qquad\!\!{}
\end{equation}
\begin{equation}
\begin{aligned}
M_6^{(g)} J_4^{(g)}\ &=\ \bigl(-144J_6^{(g+1)}+144 J_4^{(g+1)}J_2^{(g+1)}-32 J_2^{(g+1)}J_2^{(g+1)}J_2^{(g+1)} \\
&\ -\ 48(53{+}75 g{+}36 g^2)J_4^{(g+1)}+16(77{+}99 g{+}36g^2)J_2^{(g+1)}J_2^{(g+1)} \\
&\ -\ 96(261{+}1151 g{+}1320 g^2{+}432 g^3)J_2^{(g+1)} \\
&\ +\ 2304(159{+}1352 g{+}3672 g^2{+}4269 g^3 {+}2232 g^4{+}432 g^5)\bigr)\,M_3^{(g)} \\
&\ +\ \bigl(J_4^{(g+1)}-4(29{+}36g)J_2^{(g+1)}+72(27{+}139 g{+}160 g^2{+}48 g^3)\bigr)\,M_6^{(g)} \ ,
\end{aligned}
\end{equation}
with $M_6^{(g)} J_6^{(g)}$ displayed in the Appendix. 
They have the general structure predicted in~(\ref{generaltwine}).

Any maximally abelian subalgebra is generated by the center plus one arbitrary further element,
hence only one Liouville charge besides the angular Hamiltonian can exist.
There is no preferred choice among the words in $J_4$ and $J_6$, but
distinguished combinations are
\begin{equation}
\begin{aligned}
R_6\ \equiv\ M_3^\+ M_3^{\vphantom{\+}} \ &=\  
\sfrac13 J_6-\sfrac12 J_4J_2+\sfrac16 J_2J_2J_2-\sfrac16 (11{+}16 g{-}48g^2)J_4 \\[2pt]
&\ +\,\sfrac1{12}(13{+}24g{-}48g^2)J_2J_2+\sfrac13 (1{+}3g)(1{+}4g)(1{-}12g)J_2\ ,\\[4pt]
R_{12}\ \equiv\ M_6^\+ M_6^{\vphantom{\+}} \ &=\ 
-12 J_6J_6+12 \{ J_6, J_4\} J_2-\sfrac{16}{3}J_6J_2J_2J_2 +2 J_4J_4J_4-14 J_4J_4J_2J_2 \\
&\ +\,6J_4J_2J_2J_2J_2-\sfrac{2}{3}J_2J_2J_2J_2J_2J_2 + \ \textrm{lower-order terms}\ ,
\end{aligned}
\end{equation}
where the full expression can be found in the Appendix.
The corresponding intertwining relations involve a single intertwiner only and read
\begin{equation}
\begin{aligned}
M_3^{(g)} R_{6}^{(g)} &\= \bigl( R_{6}^{(g+1)} - (1{+}2g)\rho_{6}^{(g+1)} \bigr)\, M_3^{(g)} \und \\[4pt]
M_6^{(g)} R_{12}^{(g)} &\= \bigl( R_{12}^{(g+1)} - (1{+}2g)\rho_{12}^{(g+1)} \bigr)\, M_6^{(g)} 
\qquad\textrm{with} \\[4pt]
\sfrac{1}{36}\rho_{6}^{(g+1)} &\= 
-4J_4^{(g+1)}+4J_2^{(g+1)}J_2^{(g+1)}-2 (69{+}128 g{+}48 g^2) J_2^{(g+1)}+3(5{+}4g)(7{+}12g)(11{+}12g) \\[4pt]
\sfrac14\rho_{12}^{(g+1)} &\= -54\{J_6^{(g+1)},J_4^{(g+1)}\}-12 J_6^{(g+1)}J_2^{(g+1)}J_2^{(g+1)} 
+294 J_4^{(g+1)} J_4^{(g+1)} J_2^{(g+1)}  \\[4pt]
&\qquad -172  J_4^{(g+1)}J_2^{(g+1)}J_2^{(g+1)}J_2^{(g+1)} 
- 62J_2^{(g+1)} J_2^{(g+1)} J_2^{(g+1)} J_2^{(g+1)} J_2^{(g+1)}
+\ O(g)\ .
\end{aligned}
\end{equation}
Similarly, the combinations $M_3^\+ M_6$ and $M_6^\+ M_3$ intertwine with a single intertwiner
on either side of the relation (albeit not the same one).

Because the tetrahexahedral system is just two-dimensional, 
the existence of two conserved charges in addition to the Hamiltonian establishes its superintegrability.
Beyond this, one can construct, for integral values of~$g$, further conserved quantities,
\begin{equation}
Q^{(g)} \= M_*^{(g-1)}M_*^{(g-2)}\cdots M_*^{(1-g)}
\qquad\Longrightarrow\qquad
Q^{(g)} H_\Om^{(g)} \= Q^{(g)} H_\Om^{(1-g)} \= H_\Om^{(g)} Q^{(g)}\ ,
\end{equation}
where $M_*$ stands for either $M_3$ or $M_6$, like it was achieved for the full Calogero model in~\cite{CoLePl13}.
These introduce a $\Z_2$ grading, with $Q$ being odd and all previous charges being even.
Clearly, the $Q$ charges are algebraically independent of the even ones, 
but we expect that any two of them are related by some combination of even charges,
and so they represent only a single additional algebraically independent conserved quantity.
The anticommutator of any two $Q$ charges is some specific polynomial in $\{J_6,J_4,J_2\}$ whose order grows
linearly with~$g$. These and the commutators with the even charges are complicated but can be worked out in principle.

\section{Outlook}

\noindent
The model we have considered in this paper is a very special one, and our treatment profited from the
isometry $A_3\sim D_3$. It will be quite interesting to find out how the structure generalizes, 
firstly, to higher $n$ in $A_{n-1}$ and, secondly, to general Coxeter systems of rank~$r$. 
Clearly, for the angular system three objects are crucial:
\begin{itemize}
\addtolength{\itemsep}{-6pt}
\item the ring of Weyl-symmetric polynomials in $\{\cL_{ij}\}$ and its independent generators
\item maximal abelian subalgebras of the ring of Weyl-symmetric polynomials in $\{\cL_{ij}\}$
\item the elementary Weyl-antiinvariant polynomials in $\{\cL_{ij}\}$
\end{itemize}
Assuming superintegrability, we conjecture that 
the number of ring generators equals $2r{-}3$ and $r{-}1$ for the first and second item,
respectively, but one has to go to $n{=}5$ (or $r{=}4$) to really test the second assertion.
Especially useful will be the answer to the third item, as different intertwiners can be employed to
partially lift the growing degeneracy of the energy levels.

Naturally, an extension to trigonometric, hyperbolic or even elliptic angular Calogero systems
may be contemplated. Finally, it will be fascinating to introduce a $\cal PT$~deformation into the
angular system~\cite{Fring05,Fring12-rev}, 
because it will remove the codimension-one singular loci of the potential,
thereby connect the $n!$~disjoint particle sectors and also give meaning to $g{<}0$ states,
which are needed for the action of the odd $Q$~charges in the spectrum.
In fact, it was observed that the action of such kind of charges on physical states gets regularized 
when $\cal PT$ deformations are considered, like in the case of the complex trigonometric 
P\"oschI-Teller potential (which corresponds to the 3-particle case considered here)~\cite{spectral}.

%\bigskip
\newpage

\section*{Acknowledgments}
%\nopagebreak

\noindent
This work was partially supported by
the Alexander von Humboldt Foundation under grant CHL~1153844~STP,
by the Deutsche Forschungsgemeinschaft under grant LE 838/12-1.
The Centro de Estudios Cient\'{\i}ficos (CECs) is funded by the Chilean government
through the Centers of Excellence Base Financing Program of Conicyt.
This article is based upon work from COST Action MP1405 QSPACE, 
supported by COST (European Cooperation in Science and Technology).

\bigskip

\begin{appendix}

\section{Appendix: }\label{apa}

\noindent
We collect the more lengthy formulae here.
%The full form of the second intertwiner can be brought into the form
%\begin{align*}
%M_6 &\=
%\ .
%\end{align*}
%
%\newpage
%
%\noindent
The most complex elementary intertwining relation reads
\begin{align*}
M_6^{(g)} J_6^{(g)}  &\= \Big(J_6^{(g+1)}+\Gamma_{66}^{64}J_4^{(g+1)} +\Gamma_{66}^{622}J_2^{(g+1)}J_2^{(g+1)}  +\Gamma_{66}^{62} J_2^{(g+1)}+\Gamma_{66}^{60}  \Big)\,M_6^{(g)} \\ 
&\ + \Big( -144 J_6^{(g+1)}J_2^{(g+1)}-36 J_4^{(g+1)}J_4^{(g+1)}+168 J_4^{(g+1)}J_2^{(g+1)}J_2^{(g+1)} \Big. \\
&\ - 36 J_2^{(g+1)}J_2^{(g+1)}J_2^{(g+1)}J_2^{(g+1)} + \Gamma_{66}^{36} J_6^{(g+1)}+\Gamma_{66}^{342} J_4^{(g+1)}J_2^{(g+1)}+\Gamma_{66}^{3222}J_2^{(g+1)}J_2^{(g+1)}J_2^{(g+1)} \\[4pt]
&\ +\Big.\Gamma_{66}^{34}J_4^{(g+1)} + \Gamma_{66}^{322} J_2^{(g+1)}J_2^{(g+1)} +\Gamma_{66}^{34}J_4^{(g+1)}+\Gamma_{66}^{322} J_2^{(g+1)}J_2^{(g+1)}+\Gamma_{66}^{32}J_2^{(g+1)} +\Gamma_{66}^{30} \Big)\, M_3^{(g)} \ ,
\end{align*}
with the abbreviations
\begin{align*}
\Gamma_{66}^{64}&=-(107+84 g)\, ,\\[4pt]
\Gamma_{66}^{622}&= -3(23+36 g)\, ,\\[4pt]
\Gamma_{66}^{62}&=2 (4411+13438 g+12432 g^2+3168 g^3) \, , \\[4pt] 
\Gamma_{66}^{60}&=-24 (4826+33989 g+71466 g^2+64056 g^3+25056 g^4+3456 g^5)\, ,\\[4pt] 
\Gamma_{66}^{36}&=48 (266+255 g+36 g^2)\, ,\\[4pt]
\Gamma_{66}^{342}&=-48 (329+306 g+60 g^2) \, ,\\[4pt]
\Gamma_{66}^{3222}&=16 (271+245 g+48 g^2)\, , \\[4pt] 
\Gamma_{66}^{34}&=48 (4486+10765 g+8184 g^2+2160 g^3) \, ,\\[4pt] 
\Gamma_{66}^{322}&=-16(8071+22919 g+20280 g^2+5616 g^3)\, , \\[4pt] 
\Gamma_{66}^{32}&=96 (23702+151661 g+328148 g^2+317796 g^3+139680 g^4+22464 g^5) \, , \\[4pt] 
\Gamma_{66}^{30}&=-2304 (13134{+}126277 g{+}440735 g^2{+}759693 g^3{+}715926 g^4{+}372744 g^5{+}99360 g^6{+}10368 g^7) \ .
\end{align*}

\newpage 

\noindent
The full expression for the distinguished conserved charge $R_{12}\equiv M_6^\+ M_6^{\vphantom{\+}}$ is
\begin{align*} 
R_{12} &\=-12 J_6J_6+12\bigl(J_6J_4+J_6J_4\bigr)J_2-\sfrac{16}{3}J_6J_2J_2J_2
+2 J_4J_4J_4-14 J_4J_4J_2J_2+6J_4J_2J_2J_2J_2\\[4pt] 
&\ -\sfrac{2}{3}J_2J_2J_2J_2J_2J_2+\gamma_6^{64}\bigl(J_6J_4+J_6J_4\bigr)+\gamma_6^{622}J_6J_2J_2 
+ \gamma_6^{442}J_4J_4J_2+\gamma_6^{4222}J_4J_2J_2J_2\\[4pt]
&\ + \gamma_6^{22222}J_2J_2J_2J_2J_2 +\gamma_6^{62}J_6J_2+\gamma_6^{44}J_4J_4+\gamma_6^{422}J_4J_2J_2
+\gamma_6^{2222}J_2J_2J_2J_2+\gamma_{6}^{6} J_6+\gamma_{6}^{42} J_4J_2\\[4pt] 
&\ +\gamma_{6}^{222} J_2J_2J_2+\gamma_{6}^{4} J_4+\gamma_{6}^{22}J_2J_2+\gamma_{6}^{2}J_2 \ , 
\end{align*}
where the $g$-dependent coefficients are
\begin{align*}
\gamma_6^{64}&=-4 (140 + 39 g + 36 g^2) \, , \\[4pt] 
\gamma_6^{622}&=\sfrac{1120}{3} - 24 g + 96 g^2\, ,\\[4pt] 
\gamma_6^{442}&=8 (202 + 63 g + 30 g^2) \, , \\[4pt] 
\gamma_6^{4222}&=-\sfrac{8}{3}(388 + 65 g + 48 g^2) \, ,\\[4pt] 
\gamma_6^{22222}&=\sfrac{16}{3}(31 + 3 g + 3 g^2)\, ,\\[4pt]
\gamma_6^{62}&=-64 (322 + 203 g - 210 g^2 - 216 g^3) \, ,\\[4pt] 
\gamma_6^{44}&=16 (553 + 308 g - 1884 g^2 - 648 g^3)\, ,\\[4pt] 
\gamma_6^{422}&= \sfrac{8}{3} (7045 +4352 g + 2712 g^2 - 4320 g^3)\, , \\[4pt] 
\gamma_6^{2222}&=-\sfrac{8}{3}(2157 + 944 g + 48 g^2 - 1296 g^3) \, , \\[4pt] 
\gamma_6^{6}&=-96 (3940 + 1469 g - 8691 g^2 - 4584 g^3 + 5328 g^4 + 3456 g^5) \, , \\[4pt] 
\gamma_6^{42}&=16 (35466 + 18288 g - 97805 g^2 - 54456 g^3 + 56880 g^4 + 44928 g^5) \, , \\[4pt] 
\gamma_6^{222}&=-\sfrac{16}{3} (33074 + 18470 g - 94827 g^2 - 41064 g^3 + 51984 g^4 + 38016 g^5)\, , \\[4pt] 
\gamma_6^{4}&=192 (5864+ 6158 g - 57481 g^2 - 32799 g^3 + 119796 g^4 + 68112 g^5 - 60480 g^6 - 41472 g^7) \, , \\[4pt] 
\gamma_6^{22}&=-32 (19568{+}28410 g{-}165011 g^2{-}106098 g^3{+}323232 g^4{+}146592 g^5{-}138240 g^6{-}82944 g^7)\, ,\\[4pt]
\gamma_6^{2}&=-384 (448 - 2274 g - 32767 g^2 - 58215 g^3 + 52068 g^4 + 64296 g^5 - 89856 g^6 - 51840 g^7)\ . 
\end{align*}

\newpage

\noindent
Finally, the shift $\rho_{12}^{(g+1)}=\sfrac{1}{1+2g}\bigl(R_{12}^{(g+1)}-R_{12}^{(-g)}\bigr)$ takes the form
\begin{equation*}
\begin{aligned}
\sfrac14\rho_{12}^{(g+1)} &\=  -54\bigl(J_6^{(g+1)}J_4^{(g+1)}+J_4^{(g+1)}J_6^{(g+1)}\bigr) -12 J_6^{(g+1)}J_2^{(g+1)}J_2^{(g+1)} +294 J_4^{(g+1)} J_4^{(g+1)} J_2^{(g+1)} \\[4pt]
&\ -172  J_4^{(g+1)}J_2^{(g+1)}J_2^{(g+1)}J_2^{(g+1)} - 62J_2^{(g+1)} J_2^{(g+1)} J_2^{(g+1)} J_2^{(g+1)} J_2^{(g+1)} +\eta_{12}^{62}J_6^{(g+1)}J_2^{(g+1)} \\[4pt] 
&\ +\eta_{12}^{44}J_4^{(g+1)}J_4^{(g+1)}+\eta_{12}^{422}J_4^{(g+1)}J_2^{(g+1)}J_2^{(g+1)}+\eta_{12}^{2222}J_2^{(g+1)}J_2^{(g+1)}J_2^{(g+1)}J_2^{(g+1)} +\eta_{12}^{6}J_6^{(g+1)} \\[4pt] 
&\ +\eta_{12}^{42}J_4^{(g+1)}J_2^{(g+1)} +\eta_{12}^{222}J_2^{(g+1)}J_2^{(g+1)}J_2^{(g+1)}+\eta_{12}^{4}J_4^{(g+1)}+\eta_{12}^{22}J_2^{(g+1)}J_2^{(g+1)} +\eta_{12}^{2}J_2^{(g+1)}+\eta_{12}^{0} \ ,
\end{aligned}
\end{equation*}
and the nontrivial coefficients are as follows,
\begin{align*}
\eta_{12}^{62}&=36 (25 + 304 g + 144 g^2)\, ,\\[4pt]
\eta_{12}^{44}&=-3 (1739 + 4008 g + 1152 g^2)\, , \\[4pt] 
\eta_{12}^{422}&=2 (737 - 3924 g - 3168 g^2)\, , \\[4pt]
\eta_{12}^{2222}&=10013 + 23048 g + 2208 g^2 \, , \\[4pt] 
\eta_{12}^{6}&=-24 (274 + 8397 g + 27252 g^2 + 22032 g^3 + 5184 g^4)  \, , \\[4pt] 
\eta_{12}^{42}&=12 (10096 + 71365 g + 159372 g^2 + 118224 g^3 + 22464 g^4) \, , \\[4pt] 
\eta_{12}^{222}&=-4 (151692 + 632591 g + 717108 g^2 + 116208 g^3 + 19008 g^4) \, , \\[4pt] 
\eta_{12}^{4}&=-24 (82623 + 628336 g + 1923894 g^2 + 2950776 g^3 + 2270592 g^4 + 767232 g^5 + 124416 g^6) \, , \\[4pt] 
\eta_{12}^{22}&=12 (1509081 + 9095222 g + 18576756 g^2 + 13405872 g^3 + 1368576 g^4 + 442368 g^5 + 82944 g^6)\, , \\[4pt] 
\eta_{12}^{2}&=-96 (2838989{+}22731865 g{+}67596615 g^2{+}87293784 g^3{+} 37690416 g^4{-}6715008 g^5{-}1824768 g^6)\, ,\\[4pt]
\eta_{12}^{0}&=1728 (972216 + 9826815 g + 39041714 g^2 + 75013828 g^3 + 65364036 g^4 + 10127664 g^5 \\[4pt] 
&\qquad - 15564096 g^6 - 4084992 g^7) \ .
\end{align*}

\end{appendix}

\begin{sidewaysfigure}
%{\small
\begin{equation*} 
\begin{tabular}{|c|cccc|}
\hline
$\ell$ & $v_\ell^{(0)}$ & $v_\ell^{(1)}$ & $v_\ell^{(2)}$ & $v_\ell^{(3)}\vphantom{\Big|}$  \\ 
\hline
0 & 
$1$ & $\c$ & $\c^2$  & $\c^3$  \\
3 & 
$\s$ & $\c\,\s$  & $\c^2\s$ & $\c^3\s$ \\
6 & 
$\c^2{-}\s^2$ & $\c(\c^2{-}3\s^2)$ &  $\c^2(\c^2{-}5\s^2)$ & $\c^3 (\c^2 {-} 7 \s^2)$  \\
9 & 
$3\c^2\s{-}\s^3$ & $\c(\c^2\s{-}\s^3)$ & $\c^2(3\c^2\s{-}5\s^3)$ & $\c^3 (3 \c^2 \s {-} 7 \s^3)$ \\
12& 
$\c^4{-}6\c^2\s^2{+}\s^4$ & $\c(\c^4{-}10\c^2\s^2{+}5\s^4)$ & $\c^2(3\c^4{-}42\c^2\s^2{+}35\s^4)$ & $\c^3 (\c^4 {-}18 \c^2 \s^2 {+} 21 \s^4)$ \\
15& 
$5\c^4\s{-}10\c^2\s^3{+}\s^5$ & $\c(3\c^4\s{-}10\c^2\s^3{+}3\s^5)$ & $\c^2(3\c^4\s{-}14\c^2\s^3{+}7\s^5)$ & $\c^3 ( 5  \c^4 \s {-} 30 \c^2 \s^3 {+} 21 \s^5)$ \\
18& 
$\c^6{-}15\c^4\s^2{+}15\c^2\s^4{-}\s^6$ & $\c(\c^6{-}21\c^4\s^2{+}35\c^2\s^4{-}7\s^6)$ & 
$ \c^2 ( \c^6 {-} 27 \c^4 \s^2 {+} 63 \c^2 \s^4 {-} 21 \s^6)$
& $\c^3 (5 \c^6 {-} 165 \c^4 \s^2 {+} 495 \c^2 \s^4 {-} 231 \s^6) $  \\
21&
$7\c^6\s{-}35\c^4\s^3{+}21\c^2\s^5{-}\s^7$ & 
$\c (\c^6 \s{-}7 \c^4 \s^3{+}7\c^2 \s^5{-}\s^7)$ &
$\c^2 (5 \c^6 \s {-} 45 \c^4 \s^3 {+} 63 \c^2 \s^5 {-} 15 \s^7)$
& $\c^3 (5 \c^6 \s {-} 55 \c^4 \s^3 {+} 99 \c^2 \s^5 {-} 33 \s^7) $ \\
\hline
\end{tabular}
\end{equation*}
%}
\begin{equation*}
\textrm{ Table 1: \ Low-lying wave functions $v_\ell^{(g)}$ of the P\"oschl-Teller model. 
\quad Notation: \quad} 
\s \equiv \sin(3\phi) \and \c \equiv \cos(3\phi)
\end{equation*}
\end{sidewaysfigure}

%\newpage

\begin{sidewaysfigure}
%{\tiny
{\small
\begin{equation*}
\begin{tabular}{|c|cc|c|}
\hline
$\ell$ & $\ell_3$ & $\ell_4$ & $\widetilde{h}_{\ell_3,\ell_4}^{(0)}\vphantom{\Big|}$ \\ 
\hline
0 & 0 & 0 & $\{000\}$ \\
3 & 1 & 0 & $\{111\}$ \\
4 & 0 & 1 & $\{400\}-3\{220\}$ \\
6 & 2 & 0 & $\{600\}-15\{420\}+30\{222\}$ \\
7 & 1 & 1 & $3\{511\}-5\{331\}$ \\
8 & 0 & 2 & $\{800\}-28\{620\}+35\{440\}$ \\
9 & 3 & 0 & $9\{711\}-63\{531\}+70\{333\}$ \\
10& 2 & 1 & $\{1000\}-45\{820\}+42\{640\}+504\{622\}-630\{442\}$ \\
11& 1 & 2 & $5\{911\}-60\{731\}+63\{551\}$ \\
12& 4 & 0 & $36\{1200\}-2376\{1020\}+2445\{840\}+46125\{822\}+4893\{660\}-215250\{642\}+179375\{444\}$ \\
12& 0 & 3 & $101\{1200\}-6666\{1020\}+47100\{840\}+8685\{822\}-42609\{660\}-40530\{642\}+33775\{444\}$ \\
\hline
\end{tabular}
\end{equation*}
\begin{equation*}
\begin{tabular}{|c|cc|c|}
\hline
$\ell$ & $\ell_3$ & $\ell_4$ & $\widetilde{h}_{\ell_3,\ell_4}^{(1)}\vphantom{\Big|}$ \\
\hline
0 & 0 & 0 & $\{000\}$ \\
3 & 1 & 0 & $\{111\}$ \\
4 & 0 & 1 & $3\{400\}-11\{220\}$ \\
6 & 2 & 0 & $3\{600\}-39\{420\}+196\{222\}$ \\
7 & 1 & 1 & $5\{511\}-13\{331\}$   \\
8 & 0 & 2 & $\{800\}-20\{620\}+23\{440\}+12\{422\}$ \\
9 & 3 & 0 & $3\{711\}-27\{531\}+56\{333\}$  \\
10& 2 & 1 & $15\{1000\}-425\{820\}+576\{640\}+7568\{622\}-14454\{442\}$  \\
11& 1 & 2 & $35\{911\}-476\{731\}+477\{551\}+204\{533\}$ \\
12& 4 & 0 & $12\{1200\}-456\{1020\}+657\{840\}+13581\{822\}+1137\{660\}-88842\{642\}+114007\{444\}$  \\
12& 0 & 3 & $813 \{1200\}-30894\{1020\}+165652\{840\}+72131\{822\}-147943\{660\}-169702\{642\}+57527\{444\}$ \\
\hline
\end{tabular}
\end{equation*}
\begin{equation*}
\begin{tabular}{|c|cc|c|}
\hline
$\ell$ & $\ell_3$ & $\ell_4$ & $\widetilde{h}_{\ell_3,\ell_4}^{(2)}\vphantom{\Big|}$ \\
\hline
0 & 0 & 0 & $\{000\}$ \\
3 & 1 & 0 & $\{111\}$ \\
4 & 0 & 1 & $5\{400\}-19\{220\}$ \\
6 & 2 & 0 & $5\{600\}-63\{420\}+506\{222\}$ \\
7 & 1 & 1 & $\{511\}-3\{331\}$ \\
8 & 0 & 2 & $35\{800\}-644\{620\}+729\{440\}+552\{422\}$  \\
9 & 3 & 0 & $21\{711\}-207\{531\}+634\{333\}$  \\
10& 2 & 1 & $\{1000\}-25\{820\}+38\{640\}+592\{622\}-1374\{442\}$ \\
11& 1 & 2 & $21\{911\}-300\{731\}+303\{551\}+200\{533\}$ \\
12& 4 & 0 & $980\{1200\}-31752\{1020\}+51045\{840\}+1248957\{822\}+84757\{660\}-9568482 \{642\}+16884639 \{444\}$ \\
12& 0 & 3 & $411005\{1200\}-13316562\{1020\}+65753340\{840\}+36003897\{822\}-58071653\{660\}-80991402\{642\}+7379439\{444\}$ \\
\hline
\end{tabular}
\end{equation*}
}
%}
\begin{equation*}
\textrm{ Table 2: Some wave functions $v_\ell^{(g)}=r^{-\ell-6g}\De^g\widetilde{h}_\ell^{(g)}$ 
of the terahexahedric model. }
\{rst\} \ :=\ x^ry^sz^t+x^ry^tz^s+x^sy^tz^r+x^sy^rz^t+x^ty^rz^s+x^ty^sz^r
\end{equation*}
\end{sidewaysfigure}

%\bigskip
\newpage

\small{

}

\end{document}